\documentstyle[epsfig,12pt]{article}
\newcommand{\be}{\begin{eqnarray}}
\newcommand{\ee}{\end{eqnarray}}

\def\lsim{\mathrel{\rlap{\lower4pt\hbox{\hskip1pt$\sim$}}
\raise1pt\hbox{$<$}}}               
\def\gsim{\mathrel{\rlap{\lower4pt\hbox{\hskip1pt$\sim$}}
\raise1pt\hbox{$>$}}}               

\begin{document}

\rightline{\Large{Preprint RM3-TH/01-2}}

\vspace{1cm}

\begin{center}

\Large{Neutron structure function and inclusive $DIS$\\ from $^3H$ and
$^3He$ at large Bjorken-$x$\footnote{\bf To appear in Physical Review C.}}

\vspace{1cm}

\large{M.M. Sargsian$^{(1)}$, S. Simula$^{(2)}$ and M.I. Strikman$^{(3)}$}

\vspace{0.5cm}

\normalsize{
$^{(1)}$Department of Physics, Florida International University, Miami, FL
33199, USA\\
$^{(2)}$INFN, Sezione Roma III, Via della Vasca Navale 84, I-00146 Roma,
Italy\\
$^{(3)}$Department of Physics, Pennsylvania State University, University
Park, PA 16802, USA}

\end{center}

\vspace{1cm}

\begin{abstract}

\noindent A detailed study of inclusive deep inelastic scattering ($DIS$) from mirror $A = 3$ nuclei at large values of the Bjorken variable $x$ is presented. The main purpose is to estimate the theoretical uncertainties on the extraction of the neutron $DIS$ structure function from such nuclear measurements. On one hand, within models in which no modification of the bound nucleon structure functions is taken into account, we have investigated the possible uncertainties arising from: ~ i) charge symmetry breaking terms in the nucleon-nucleon interaction, ~ ii) finite $Q^2$ effects neglected in the Bjorken limit, ~ iii) the role of different prescriptions for the nucleon Spectral Function normalization providing baryon number conservation, and ~ iv) the differences between the virtual nucleon and light cone formalisms. Although these effects have been not yet considered in existing analyses, our conclusion is that all these effects cancel at the level of $\approx 1\%$ for $x \lsim 0.75$ in overall agreement with previous findings. On the other hand we have considered several models in which the modification of the bound nucleon structure functions is accounted for to describe the $EMC$ effect in $DIS$ scattering from nuclei. It turns out that within these models the cancellation of nuclear effects is expected to occur only at a level of $\approx 3 \%$, leading to an accuracy of $\approx 12 \%$ in the  extraction of the neutron to proton structure function ratio at $x \approx 0.7 \div 0.8$. Another consequence of considering a broad range of models of the $EMC$ effect is that the previously suggested  iteration procedure does not improve  the accuracy of the extraction of the neutron to proton structure function ratio.

\end{abstract}

\vspace{1cm}


\vspace{0.25cm}


\newpage

\pagestyle{plain}

\section{Introduction}

\indent The investigation of deep inelastic scattering ($DIS$) of leptons off the nucleon is an important tool to get fundamental information on the structure of quark distributions in the nucleon. In the past years several experiments have been performed in order to study the region of small values of the Bjorken variable $x \equiv Q^2 / 2M \nu$, which is dominated by sea quarks and gluons. Recently, experiments at $HERA$ have pushed the measurements at large $x$ to a new high-$Q^2$ domain, while dijet measurements at Tevatron have reached the kinematics where the knowledge of the quark distributions in the nucleon at $x \gsim 0.6$ becomes important (for a recent discussion and references see \cite{Yang}).

\indent One of the major uncertainty in large-$x$ studies comes from a poor knowledge of the $d$-quark distribution in the nucleon. The reason is that it is very difficult to extract $d(x, Q^2)$ from measurements off hydrogen targets: it enters as a correction in case of inclusive electron scattering off the proton at $Q^2 \ll M_W^2$ (where $M_W$ is the mass of the $W$-boson), while the measurements using semi-inclusive $e + p \to e + \pi^{\pm} + X$ and large-$Q^2$ $e^{\pm} + p \to \nu (\bar{\nu}) + X$ scatterings do not have reached a sufficient degree of statistical accuracy yet. As a result, one has to rely on the extraction of $d(x, Q^2)$ from the data involving $DIS$ off the deuteron.

\indent On the theoretical side the predictions for the behavior of the ratio $d(x, Q^2) / u(x, Q^2)$ at $x \to 1$ vary very significantly. The deviations from the $SU(6)$ value $d(x, Q^2) / u(x, Q^2) = 0.5$ could come either from non-perturbative effects, which could lead, as suggested firstly by Feynman \cite{Feynman}, to $d(x, Q^2) / u(x, Q^2) \to 0$ (corresponding to $F_2^n(x, Q^2) / F_2^p(x, Q^2) \to 1 / 4$) at $x \to 1$ \cite{Isgur} or from the hard scattering (perturbative $QCD$) mechanism, yielding $d(x, Q^2) / u(x,Q^2) \to 1/5$  (corresponding to $F_2^n(x, Q^2) / F_2^p(x, Q^2) \to 3 / 7 \simeq 0.43$) at $x \to 1$ \cite{pQCD}. Thus experimental data on $F_2^n(x, Q^2) / F_2^p(x, Q^2)$ at large $x$ have a high degree of theoretical significance.

\indent The first extraction of the ratio $F_2^n(x, Q^2) / F_2^p(x, Q^2)$ from the $SLAC$ $DIS$ $p(e, e')X$ and $D(e, e')X$ data (see \cite{Bodek}(a) and references therein) used the West procedure \cite{West}, which is based on a covariant electron-deuteron scattering formalism with the interacting nucleon off-mass-shell and the spectator nucleon on-mass-shell. Such a procedure leads to the so-called West correction $\sigma_{tot}(\gamma^* + d) / (\sigma_{tot}(\gamma^* + p) + \sigma_{tot}(\gamma^* + n)) < 1$ in the impulse approximation (the numerical value of this ratio is around $0.980 \div 0.985$). The application of the West procedure has provided a limiting value $(F_2^n / F_2^p)_{\left| x \to 1 \right.} \approx 1 / 4$, which has been adopted in most of the global fits of parton distribution functions ($PDF$'s) (see, e.g., \cite{DATA,SLAC,CTEQ,GRV}). Later \cite{FS76west,LP78} it was pointed out that the West correction leads to a violation of the Gross-Llewellyn Smith (baryon charge) sum rule, because of the neglect of relativistic corrections in the normalization of the deuteron wave function in Ref. \cite{West}. Furthermore it was pointed out in \cite{FS76west} that modeling the deuteron wave function with one on-mass-shell- and one off-mass-shell nucleon without taking into account other degrees of freedom unavoidably leads to the violation of the energy-momentum sum rule, which expresses the requirement that the sum of the light cone fractions carried by all partons adds up to $1$. Then an alternative light-cone ($LC$) formalism was suggested \cite{FS76lc} which satisfies both the baryon charge and the momentum sum rules. Both the Virtual Nucleon Convolution ($VNC$) model \cite{FS76west} and the $LC$ formalism \cite{FS76lc} lead to an enhancement of $F_2^D(x, Q^2)$ at large $x$ as compared to the $SLAC$ procedure and hence to a further decrease of the extracted value of the $n / p$ ratio, $F_2^n / F_2^p$, at large $x$.

\indent The discovery of the $EMC$ effect \cite{EMCexp} has clearly indicated gross deviations of the $F_2^A / F_2^D$ ratio from the predictions based on the Fermi motion approximation in the kinematical regions at large $x$ relevant for the extraction of $F_2^n / F_2^p$ from the deuteron inclusive data. This  led immediately to the conclusion \cite{FS85} that the value of $F_2^n / F_2^p$ extracted  by $SLAC$ was underestimated. Therefore an approximate procedure, which was argued to depend only marginally on the details of the $EMC$ effect \cite{FS85,FS,S92}, was suggested for $x \lsim 0.7$. The use of this procedure by the $SLAC$ experimental group \cite{BDR} has confirmed the conclusion of Ref. \cite{FS85}, finding that the value of $F_2^n / F_2^p$ at $x \approx 0.7$ may be much closer to the  $pQCD$ asymptotic value of $3 / 7 \simeq 0.43$.

\indent Over the years a number of further studies of deuteron structure functions have been performed using the $VNC$ model \cite{OFF-SHELL} and most of them have adopted the normalization of the deuteron wave function of Ref. \cite{FS76west}. Some of these studies have included also pion degrees of freedom to fix the momentum sum rule problem. Other studies have considered off-mass-shell effects in the structure function of the interacting virtual nucleon. Within the latter the issue of the extraction of the ratio $F_2^n / F_2^p$ was analyzed in Ref. \cite{MT96}, where the extracted values for the $n / p$ ratio at $x \lsim 0.7$ turned out to be very similar to the findings of Refs. \cite{BDR,S92}.

\indent It should be also mentioned that quite recent analyses \cite{Yang,Ricco_NPB} of leading and higher twists in proton and deuteron $DIS$ data have found that the latter are consistent with a significant enhancement of the $d$-quark distribution at large $x$ with respect to the standard $PDF$ behavior of $d/u \rightarrow 0$.

\indent The realization that the extraction of the large-$x$ $n / p$ ratio from deuteron $DIS$ data is inherently model dependent has led to the suggestion of two new strategies. One is the use of the tagged semi-inclusive processes off the deuteron \cite{Strikman,SIM96}, which require the detection of a low momentum proton ($p \lsim 150 MeV/c$). The other one is the determination of the $DIS$ structure functions of mirror $A = 3$ nuclei \cite{Wally,Pace}. In the former one can {\em tag} the momentum of the struck neutron by detecting the slow recoiling proton; in this way it is possible to select initial deuteron configurations in which the two nucleons are far apart, so that the struck nucleon can be considered as free. In principle, one can use here an analog of the Chew-Low  procedure for the study of scattering off a pion \cite{Chew} and extrapolate the cross section to the neutron pole. The neutron structure function can then be extracted directly from the semi-inclusive deuteron cross section without significant nuclear model dependence \cite{Strikman,SIM96}. In the latter one tries to exploit the mirror symmetry of $A = 3$ nuclei; in other words, thanks to charge symmetry, one expects that the magnitude of the $EMC$ effect in $^3He$ and $^3H$   \be
      {\cal{R}}_{EMC}^A(x, Q^2) \equiv {F_2^A(x, Q^2) \over F_2^D(x, Q^2)} ~
      {F_2^p(x, Q^2) + F_2^n(x, Q^2) \over ZF_2^p(x, Q^2) + N F_2^n(x, Q^2)}
      \label{eq:REMC}
 \ee
is very similar and hence the so called super-ratio \cite{Wally}
 \be
      {\cal{SR}}_{EMC}(x, Q^2) \equiv {{\cal{R}}_{EMC}^{^3He}(x, Q^2) \over 
      {\cal{R}}_{EMC}^{^3H}(x, Q^2)} = {F_2^{^3He}(x, Q^2) \over 
      F_2^{^3H}(x, Q^2)} ~ {2F_2^n(x, Q^2) + F_2^p(x, Q^2) \over 2F_2^p(x, 
      Q^2) + F_2^n(x, Q^2)} ~,
      \label{eq:superatio}
 \ee
should be very close to unity regardless of the size of the $EMC$ ratios itself \cite{Wally,Pace}. If this is true, the $n / p$ ratio could be extracted directly from the ratio of the measurements of the $^3He$ to $^3H$ $DIS$ structure functions without significant nuclear modifications. However, it should be pointed out that, even if charge symmetry were exact, the motion of protons and neutrons in a non isosinglet nucleus (let's say $^3He$) is somewhat different due to the spin-flavor dependence of the nuclear force.

\indent The aim of this paper is to perform explicit calculations of the $EMC$ effect for both $^3He$ and $^3H$ targets, taking properly into account the motion of protons and neutrons in mirror $A =3$ nuclei. We explore in greater details the $VNC$ model used in \cite{Wally,Pace} in order to analyze the effects of: ~ i) charge symmetry breaking terms in the nucleon-nucleon ($NN$) interaction; ~ ii) finite $Q^2$ effects in the impulse approximation; ~ iii) the role of different prescriptions for the nucleon Spectral Function  normalization providing baryon number conservation; and iv) the role of different $PDF$ sets. Additionally we compare the predictions of the $VNC$ model and the $LC$ formalism in the approximation where no bound nucleon modification is taken into account. It will be shown that the inclusion of these additional effects leaves the super-ratio (\ref{eq:superatio}) close to unity within $1 \%$ only for $x \lsim 0.75$, confirming therefore the findings of Refs. \cite{Wally} and \cite{Pace}, where deviations of the order of $2 \%$ and $1 \%$ were found, respectively.

\indent However, it is well known that the $VNC$ model underestimates significantly the $EMC$ effect at large $x$. Also, if the VNC model is adjusted to satisfy the momentum sum rule by adding pionic degrees of freedom, it leads to a significant enhancement of the $\bar{q}_A / \bar{q}_N$ ratio at $x \gsim 0.1$, where a suppression is observed experimentally \cite{DY}. Moreover, the $VNC$ model is just one of the many models of the $EMC$ effect. Similarly the $LC$ formalism without bound nucleon modifications strongly disagrees with data at large $x$. Hence, to provide a more conservative estimate of the possible range of deviations of the super-ratio from unity we will also investigate carefully various models of the $EMC$ effect which interpret this effect as due to modification of the wave function of either individual nucleons or two nucleon correlations. We will show that the cancellation of the nuclear effects in the super-ratio (\ref{eq:superatio}) within the broad range of the models considered occurs only at the level of $\approx 3 \%$,  restricting significantly (up to $\approx 12 \%$) the accuracy of the extraction of the free $n / p$ ratio from the ratio of the measurements of the $^3He$ to $^3H$ $DIS$ structure functions.

\indent In this work we will not address all the EMC models predicting possible deviations from the convolution formula at large $x$. We feel however that it's worthwhile to mention at least few of them. An important issue in modeling the EMC effect is the possible role of final state interaction effects even in the Bjorken limit. Though these effects are absent if the scattering process is formulated directly in terms of parton degrees of freedom, the final state interactions may be present in the case of a two stage descriptions, where the nucleus is described as a system of hadrons and next the scattering off the parton constituents of the hadrons is considered. However, it is very difficult to obtain safe estimates of such an effect, and therefore we have not included it in the present work. Another issue is the polarizability of the nucleon into $\Delta(1232)$ components, which contributes to three-nucleon forces adding a $\simeq 10\%$ correction to the binding energy of the three-nucleon system. Consequently, it may be possible that at large $x$ one is not measuring only the nucleon structure function. Note here that the interference among the scattering off $\Delta$-isobar and nucleon is known to be relevant for the description of the polarized $A = 3$ structure functions \cite{g1}. The role of the $\Delta(1232)$ component effects in the problem of extraction of the $F_2^n / F_2^p$ ratio deserves a special study, which is beyond the scope of this paper.

\indent The plan of this paper is as follows. In Section 2 the formalism needed to evaluate the nuclear structure functions $F_2^A(x, Q^2)$  within the $VNC$ model and the $LC$ approximation is presented. All the necessary inputs for a realistic estimate of the basic nuclear ingredient, namely the invariant nucleon Spectral Function and its proper normalization, are discussed. In Section 3 both the nuclear $EMC$ effect and the super-ratio in mirror $A = 3$ nuclei are evaluated adopting the $VNC$ model and the $LC$ approximation, assuming also no modification of the bound nucleon structure functions. Section 4 is devoted to estimate the deviations of the super-ratio ${\cal{SR}}_{EMC}(x, Q^2)$ from unity in several models of the $EMC$ effect, in which modifications of the bound nucleon structure functions are considered. The issue of the extraction of the $n / p$ ratio from the measurement of the ratio of the mirror $A = 3$ structure functions is fully analyzed in Section 5. Our main conclusions are then summarized in Section 6.

\section{Basic Formalism for Inclusive $DIS$ from Nuclei}

\indent There exist a number of treatments in the literature. However some of them do large $Q^2$ approximations right away, do not specify completely a prescription for treating off-mass-shell effects in the amplitude of virtual photon-nucleon interaction, etc.  Hence we find it necessary in this Section to rederive the basic formulae needed for the evaluation of the nuclear structure function $F_2^A(x, Q^2)$ within the $VNC$ model and the $LC$ approach at finite $Q^2$.

\noindent In both cases no modification of the bound nucleon structure functions will be considered. We will refer to these approximations as convolution approximations.

\subsection{Virtual Nucleon Convolution Model}

\indent The cross section for the inclusive $A(e, e')X$ reaction can be written in the following general form 
 \be
      {d\sigma \over dE'_e d\Omega'_e} = {E'_e \over E_e} ~ {\alpha^2 \over 
      q^4} ~ \eta_{\mu \nu} ~ W_A^{\mu \nu},
      \label{sigma}
 \ee
where $\eta_{\mu \nu} \equiv {1 \over 2} \mbox{Tr}(\hat{k}_2 \gamma_{\mu} \hat{k}_1 \gamma_{\nu})$ is the leptonic tensor, $k_1 \equiv (E_e, \vec{k}_1)$ and $k_2 \equiv (E'_e, \vec{k}_2)$ are the four-momenta of the incident and scattered electrons, respectively, and $W_A^{\mu \nu}$ is the electromagnetic tensor of the target, viz. 
 \be
      W_A^{\mu \nu} = \sum_{spin, X} \langle A \mid J_A^{\mu}(q) \mid X 
      \rangle \langle X \mid J_A^{\mu \dag}(0) \mid A \rangle.
      \label{Wtens}
 \ee

\indent Within the covariant impulse approximation one assumes that the virtual photon interacts with a virtual nucleon and the final hadronic state $X$ consists of the product of inelastic $\gamma^*N$ interaction and the recoil ($A - 1$)-nucleon system. Based on the Feynman diagram analysis of this scattering for the electromagnetic nuclear tensor \cite{FS87} one obtains: 
 \be
      W_A^{\mu \nu} = \sum_N \int d^4p ~ S_N(p) ~ W_N^{\mu \nu}.
      \label{IA}
 \ee
Here the invariant nucleon Spectral Function in the nucleus is defined as:
 \be 
      S_N(p) = \int d[p_{A-1}] ~ \Gamma^2(p, [p_{A-1}]),
      \label{spectral}
 \ee
where $p$ is the momentum of the virtual nucleon, $[p_{A - 1}]$ denotes internal variables of the residual on-mass-shell ($A - 1$)-nucleon system and $\Gamma(p, p_{A - 1})$ is the covariant $A \to N, (A - 1)$ vertex function combined with the propagator of the virtual nucleon. Based on the requirement of baryonic number conservation \cite{FS87} the nucleon Spectral Function is normalized as follows: 
 \be
      \int d^4 p ~ A {p_0 - p_z \over M_A} ~ S_N(p) = \int d^4 p ~ {A p_0 
      \over M_A} ~ S_N(p) = 1.
      \label{norm}
 \ee 
To proceed further, we express the  electromagnetic tensor through the two invariant structure functions $W_1$ and $W_2$:
 \be
      W_j^{\mu \nu} & = & - W_1^j(p_j \cdot q, Q^2) \left( g^{\mu \nu} - 
      {q^{\mu} q^{\nu} \over q^2} \right) \nonumber \\
      & + & {W_2^j(p_j \cdot q, Q^2) \over M_j^2} \left( p_j^{\mu} - q^{\mu}
      {p_j \cdot q \over q^2} \right) \left( p_j^{\nu} - q^{\nu} {p_j \cdot 
      q \over q^2} \right),
      \label{tensor}
 \ee
where $j = A, N$. Multiplying the left and right sides of Eq. (\ref{IA}) by $\tilde{k}_1^{\mu} \equiv k_1^{\mu} -  q^{\mu} ~ k_{1-} / q_-$ (see e.g. \cite{FS,FSS90})\footnote{Note that $\tilde{k}_1^{\mu}$ automatically fulfills the current conservation $q \cdot J_j = 0$ and $\tilde k_1^- = 0$.}), where $k_{1-} = \epsilon_1 - k_{1z}$ and $q_- = q_0 - q_z$, and considering the limit of $\epsilon_1, k_{1} \to \infty$ with both $Q^2$ and $q_0$ fixed, one obtains
 \be
      W_2^A(Q^2, \nu) = \sum_N \int d^4p ~ S(p) ~ W_2^N(Q^2, \tilde{w}) 
      \left[ {1 \over M^2} (1 + \mbox{cos}\delta)^2 (p_- + q_- {M \nu' \over
      Q^2})^2 + {p_{\perp}^2 \over 2 M^2} \mbox{sin}^2\delta \right],
      \label{W2IA} 
 \ee
where $\mbox{sin}\delta = \sqrt{Q^2} / |\vec{q}|$, $p_- = p_0 - p_z$, $\nu' = p \cdot q / M$ and $\tilde w^2 = (p+q)^2$. Because of the off-shellness of the interacting nucleon one has $p^2\ne M^2$. Contracting Eq. (\ref{IA}) with the unit vector $n^{\mu} = (0,0,0,n_y)$ one has 
 \be
      W_1^A(Q^2,\nu) = \sum_N \int d^4p ~ S_N(p) ~ \left\{ W_1^N(Q^2, 
      \tilde{w}) + {p_{\perp}^2 \over 2 M^2} W_2^N(Q^2, \tilde{w}) 
      \right\}.
      \label{W1IA}
 \ee
\indent The inclusive cross section (\ref{sigma}) can be expressed through the structure function $W_2^A$ and $W_1^A$ in a standard way:
 \be
      {d\sigma \over dE_{e'} d\Omega_{e'}} = \sigma_{Mott} \left\{ 
      W_2^A(Q^2, \nu) + 2 \mbox{tan}^2({\theta_{e} \over 2}) ~ W_1^A(Q^2, 
      \nu) \right\}. 
      \label{sigmaw}
 \ee
In case of $DIS$ one introduces the usual scaling functions:
 \be
       F_1^j & = & M W_1^j , \nonumber \\
       F_2^j & = & \nu W _2^j,
       \label{ffun}
 \ee
where $j = A, N$. Using Eq. (\ref{ffun}) in Eqs. (\ref{W2IA},\ref{W1IA},\ref{sigmaw}) one gets:
 \be
      {d\sigma \over dE_{e'} d\Omega_{e'}} = \sigma_{Mott} {1 \over \nu} 
      \left\{ F_2^A(x, Q^2) + {2\nu \over M} \mbox{tan}^2({\theta_{e} \over 
      2}) F_1^A(x, Q^2) \right\},
      \label{sigmaF}
 \ee
where $x={Q^2\over 2M\nu}$ and 
 \be
    \label{F1IA}
      F_1^A(x,Q^2) & = & \sum_N \int d^4p ~ S_N(p) ~ \left\{ 
      F_1^N(\tilde{x}, Q^2) + {p_{\perp}^2 \over 2 M\nu'} F_2^N(\tilde{x}, 
      Q^2) \right\}, \\
      \label{F2IA}
      F_2^A(x,Q^2) & = & \sum_N \int d^4p ~ S_N(p) ~ F_2^N(\tilde{x}, Q^2) 
      {\nu \over {\tilde \nu}} \left[ {1 \over M^2} (1 + \mbox{cos}\delta)^2
      (p_- + q_- {M\nu' \over Q^2})^2 \right. \nonumber \\
      & & \left. + {p_{\perp}^2 \over 2 M^2} \mbox{sin}^2\delta \right],
 \ee
where $\tilde{\nu} = (\tilde{w}^2 + Q^2 - M^2) / 2M = \nu^{\prime} + (p^2 - M^2) / 2M$ and  $\tilde{x} = Q^2 / 2 M \tilde{\nu}$. Note that with such a definition of the argument of $F_j(\tilde{x}, Q^2)$ it is ensured that the cross section is vanishing below the threshold for the $e D \rightarrow e' p n$ reaction.

\subsubsection{The Nuclear Structure Function $F_2^A(x, Q^2)$} \label{NSF}

\indent In this subsection we will discuss the $DIS$ structure function $F_2^A(x, Q^2)$ in more detail. Let's introduce the scaling variables
 \be
      z & = & {A p_- \over M_A}  ~ , \nonumber \\
      \alpha_q & = & {A q_- \over M_A},
      \label{alpha}
 \ee
and make use of the identity $d^4p = {1 \over 2} dp_+d p_- d^2p_{\perp}$; then, Eq. (\ref{F2IA}) can be written as
 \be
      F_2^A(x, Q^2) & = & {1 \over 2} \sum_{N} \int dp_+d p_- d^2p_{\perp}
      dz ~ S_N(p) ~ F_2^N(\tilde{x}, Q^2) {\nu \over \tilde{\nu}} \delta(z - 
      {A p_- \over M_A}) \nonumber \\
      & \cdot & \left[ ({M_A \over AM})^2 (1 + \mbox{cos}\delta)^2 (z + 
      \alpha_q {M \nu'\over Q^2})^2 + {p_{\perp}^2 \over 2 M^2} 
      \mbox{sin}^2\delta \right].
      \label{F2IAB} 
 \ee
The integration over $dp_-$ can be taken automatically, while the integration over $p_+$, which describes the virtuality of the interacting nucleon, requires the knowledge of the invariant nucleon Spectral Function. One can proceed however by observing that the virtuality of the interacting nucleon depends on the structure of the recoil ($A - 1$)-nucleon system. Namely for the case of two-body break-up the invariant Spectral Function contains the $\delta(p_+ - p_{+0})$ function with 
 \be
      p_{+0} = M_A - {(M_{A - 1}^f)^2 + p_{\perp}^2 \over (A - z) M_A / A},
      \label{2bb}
 \ee
where $M_{A - 1}^f$ is the mass of the recoiling $A-1$ nucleus. In case of the excitation of the recoil nuclear system into its continuum, one can use the observation, based on the multinucleon correlation model \cite{2NC}, that for different ranges of $z$ the dominant value of $p_+$ in $S_N(p)$ depends on whether the interacting nucleon is in the nuclear mean field or in $2N$, $3N$, etc. correlations. Based on this model we can estimate the integrand in Eq. (\ref{F2IAB}) as
  \be
      && <p_+> = M_A - {M_{A - 1}^2 + p_{\perp}^2 \over (A - z) M_A / A}  \ 
      \ \mbox{at} \ \ z \le 1.2 \div 1.3 \nonumber \\
      && <p_+> = M_A - M_{A - 2}- {M^2 + p_{\perp}^2 \over (2 - z) M_A / A}
      \ \ \mbox{at} \ \ z > 1.2 \div 1.3 \ (2N \ correlations) \nonumber \\
      && <p_+> = M_A - M_{A - 3}- {(2M)^2 + p_{\perp}^2 \over (3 - z) M_A / 
      A} \ \ \mbox{at} \ \ z > 1.7 \div 1.8 \ (3N \ correlations)
      \nonumber \\
      &&\mbox{...}
      \label{3bb}
 \ee
Using these approximations we can now integrate Eq. (\ref{F2IAB}) over $p_+$ arriving at:
 \be
      F_2^A(x, Q^2) = && \sum_N \int dz d^2p_{\perp} ~ \rho_N(z, p_{\perp}) 
      ~ F_2^N(<\tilde{x}>, Q^2) {\nu \over <\tilde{\nu}>} \nonumber \\
      \cdot && \left[ ({M_A \over AM})^2 (1 + \mbox{cos} \delta)^2 (z + 
      \alpha_q {M <\nu'> \over Q^2})^2 + {p_{\perp}^2 \over 2 M^2} 
      \mbox{sin}^2\delta \right], 
      \label{F2IABC} 
 \ee
where 
 \be
      <\tilde{x}> & = & {Q^2 \over 2M <\tilde{\nu}>}, \nonumber \\
      <\tilde{\nu}> & = & {w^2 + Q^2 - M^2 \over 2M}, \nonumber \\ 
      w^2 & = & Q^2 + {1 \over 2}{M_A \over A}(p_+ \alpha_q + z q_+) + {M_A 
      \over A} p_+z - p_{\perp}^2, \nonumber \\
      <\nu'> & = &  {1 \over 2M} (p_+ q_- + p_- q_+) = {M_A \over AM} \left[
      p_+ \alpha_q + q_+ z \right],
      \label{pplus}
 \ee
and $p_+$ here defined according to Eqs. (\ref{2bb}, \ref{3bb}). In Eq. (\ref{F2IABC}) $\rho_N(z, p_{\perp})$ is the one-body density function in the nucleus, defined as:
 \be 
      \rho_N(z, p_{\perp}) & = & {1 \over 2} \int dp_- dp_+ ~ S_N(p_0, p_z, 
      p_{\perp}) \delta(z - {p_0 - p_z \over M_A/A}) \nonumber \\
      & = & {M_A \over A} \int dp_0 ~ S_N(p_0, p_0 - z {M_A \over A}, 
      p_{\perp}).
      \label{rho}
\ee

\subsubsection{The Bjorken Limit}

Equation (\ref{F2IABC}) allows to calculate the inelastic $A(e,e')x$ reaction in a wide range of values of $Q^2$, i.e. large enough that the condition for the closure over final hadronic states is achieved and the impulse approximation is valid. Additionally, in DIS the range of the Bjorken $x$ should correspond to the valence region ($x > 0.2 \div 0.3$) where shadowing effects are negligible. In the Bjorken limit, where $Q^2, q \to \infty$ and $x$ is kept fixed, Eq. (\ref{F2IAB}) transforms to the usual convolution formula used by many authors \cite{Wally,FS87,FS81,EMC,CL}: 
 \be
      F_2^A(x, Q^2) = \sum_{N = 1}^A \int_x^A dz ~ z ~ f^N(z) ~ F_2^N({x 
      \over z},Q^2),
      \label{eq:convolution}
\ee
where $f^N(z)$ is the nucleon light-cone momentum distribution in the nucleus
 \be
      f^N(z) = \int d^2p_\perp ~ \rho_N(z, p_{\perp}),
      \label{nlcd}
 \ee
with the baryon charge normalization  condition given by Eq. (\ref{normrho}).

\indent Introducing the compact notation $f^N \otimes F_2^N$ to indicate the convolution (\ref{eq:convolution}) and assuming {\em exact} nuclear charge symmetry, the $^3He$ and $^3H$ $DIS$ structure functions can be written as  \be
      F_2^{^3He} & = & S \otimes (2F_2^p + F_2^n) + D \otimes (2F_2^p - 
      F_2^n), \nonumber \\
      F_2^{^3H} & = & S \otimes (2F_2^n + F_2^p) + D \otimes (2F_2^n - 
      F_2^p),
     \label{eq:mirror}
 \ee
where
 \be
    S(z) & \equiv & {f^p(z) + f^n(z) \over 2}, \nonumber \\
    D(z) & \equiv & {f^p(z) - f^n(z) \over 2},
    \label{eq:SD}
 \ee
with $f^{p(n)}(z)$ being the light-cone momentum distribution of proton (neutron) in $^3He$. If $D(z) \simeq 0$ [i.e. $f^p(z) \simeq f^n(z)$], then it is reasonable to expect that the $EMC$ ratios (\ref{eq:REMC}) in $^3He$ and $^3H$ are quite close each other, so that the super-ratio (\ref{eq:superatio}) is close to unity, as observed in Ref. \cite{Wally}. However, as it will be illustrated in detail in the next subsection, the spin-flavor dependence of the nuclear force (even without any charge-symmetry and charge-independence breaking terms) yield $f^p(z) \neq f^n(z)$. Therefore, when $D(z) \neq 0$, the difference in the proton and neutron structure function [leading to $2F_2^p - F_2^n \neq 2F_2^n - F_2^p$ in Eq. (\ref{eq:mirror})] can give rise to $R_{EMC}^{^3He} \neq R_{EMC}^{^3H}$ and correspondingly to deviations of the super-ratio (\ref{eq:superatio}) from unity depending on the size of the $EMC$ effect itself. {\em It is important to note that the nuclear charge symmetry will not  limit such deviations.}

\subsection{Nuclear Density Function and $LC$ Momentum Distribution in the $VNC$ Model}

Now we further analyze the one-body density function $\rho_N(z, p_{\perp})$ and the light-cone momentum distribution $f^N(z)$. From Eqs. (\ref{norm}) and (\ref{rho}) one obtains the following normalization for the one-body density function \cite{FS87}: 
 \be
      \int d^2p_{\perp} \int_0^A dz ~ z ~ \rho_N(z, p_{\perp}) = \int_0^A dz
      ~ z ~ f^N(z) = 1.
      \label{normrho}
 \ee
To construct the one-body density function and subsequently the light-cone momentum distribution we have to relate the invariant Spectral Function $S_N(p)$ to the non-relativistic Spectral Function $P^N(p, E)$, which represents the joint probability to find in the nucleus a nucleon with three-momentum $p = |\vec{p}|$ and removal energy $E$. Since the latter is defined as $E \equiv E_A - E_{A-1} + E_{A-1}^*$ [$E_{A-1}^*$ being the (positive) excitation energy of the system with ($A-1$) nucleons measured with respect to its ground-state, and $E_A$ ($E_{A-1}$) the binding energy of the nucleus $A$ ($A-1$)], the nucleon Spectral Function also represents the probability that, after a nucleon with momentum $p$ is removed from the target, the residual ($A-1$)-nucleon system is left with excitation energy $E_{A-1}^*$.

\indent Since the (non-relativistic) nucleon Spectral Function $P^N(p, E)$ is normalized as
  \be
       4\pi  \int_{E_{min}}^{\infty} dE \int_0^{\infty} dp ~ p^2 ~ P^N(p, E)
       = 1,
       \label{eq:norm}
 \ee
where $E_{min} \equiv E_A - E_{A-1}$ is the minimum value of the removal energy, one has some ambiguity in the relation between $S_N(p)$ and $P^N(p, E)$. Two ansatz were suggested to relate $S_N(p)$ and $P^N(p, E)$ which can be considered to represent two extremes. In one \cite{FS,FS87,FSS90} it is assumed that 
 \be 
      S_N(p) = {M_A \over Ap_0} \cdot P^N(p, E).
      \label{fsnorm}
 \ee
In this case the non-relativistic transition from $S_N(p)$ to $P^N(p, E)$ is straightforward, since in this limit one has $(M_A / Ap_0) \to \approx 1$, and hence the renormalization is the smallest for small nucleon momenta. Another ansatz assumes \cite{CL} that renormalization is momentum independent so that 
 \be
      S_N(p) \approx C_N \cdot P_N(p, E),
      \label{clnorm}
\ee
where $C_N$ can be found from the requirement given by Eq. (\ref{norm}). In this prescription $f^N(z)$ reads explicitly as
 \be
       f^N(z) = 2\pi M C_N \int_{E_{min}}^{\infty} dE \int_{p_{min}(z,
       E)}^{\infty} dp ~ p ~ P^N(p, E) {M \over \sqrt{M^2 + p^2}},
      \label{eq:fN}
 \ee
where $p_{min}(z, E)$ is  given by
 \be
       p_{min}(z, E) = {1 \over 2} {(M_A - Mz)^2 - (M_{A-1}^f)^2 \over M_A -
       Mz},
      \label{eq:kmin}
 \ee
with $M_A = A \cdot M + E_A$ and $M_{A-1}^f = M_{A-1} + E - E_{min}$. Note that the normalization factor $C_N$ can be different for protons and neutrons, in line with the normalization factor appearing in Eq. (\ref{fsnorm}).

\indent For a generic nucleus with  $A > 2$ the nucleon Spectral Function can be written as (cf., e.g., \cite{2NC,CL}) 
 \be
       P^N(p, E) = P_0^N(p, E) + P_1^N(p, E),
       \label{eq:PpE}
 \ee
where $P_0^N(p, E)$ includes the contributions of all the final states belonging to the discrete spectrum of the ($A - 1$)-nucleon system (basically its ground and one-hole states), while $P_1^N(p, E)$ corresponds to more complex final configurations (i.e. the final states of the continuum spectrum of the ($A - 1$)-nucleon system), which are mainly 1p-2h states arising from the 2p-2h excitations generated in the target ground state by short-range and tensor $NN$ correlations. In what follows we will refer to $P_0^N$ and $P_1^N$ as the ground and correlated parts of the nucleon Spectral Function, respectively.

\indent The nucleon momentum distribution $n^N(p)$ can be simply obtained from the nucleon Spectral Function by integrating over the removal energy; thus, Eq. (\ref{eq:PpE}) implies that $n^N(p)$ can be written as the sum of two components related to the ground and correlated parts of $P^N(p, E)$, respectively, viz. 
 \be
       n^N(p) \equiv \int_{E_{min}}^{\infty} dE ~ P^N(p, E) = n_0^N(p) +
       n_1^N(p).
       \label{eq:np}
 \ee
Useful parameterizations of the results of many-body calculations of $n^N(p)$, available for few-nucleon systems, complex nuclei and nuclear matter, as well as its decomposition (\ref{eq:np}) into $n_0^N(p)$ and $n_1^N(p)$ can be read off from Ref. \cite{2NC}(b).

\indent As it is well known, the calculation of $P^N(p, E)$ for $A > 2$ requires the knowledge of a complete set of wave functions for ($A - 1$) interacting nucleons. Thus, since the latter ones should be obtained from many-body calculations using realistic models of the $NN$ interaction, the evaluation of $P^N(p, E)$ represents a formidable task. In case of $^3He$ the nucleon Spectral Function has been obtained using three-body Faddeev \cite{Sauer} or variational \cite{CPS} wave functions, whereas for $A = \infty$ the evaluation of $P^N(p, E)$ has been performed using the orthogonal correlated basis approach \cite{BFF} and perturbation expansions of the one-nucleon propagator \cite{propagator}. Since in this work we are interested in the evaluation of Eqs. (\ref{F2IABC},\ref{eq:convolution}) for mirror $A = 3$ nuclei as well as for $A > 3$ nuclei, in what follows we will adopt the Spectral Function model of  Ref. \cite{2NC}, which was developed for any value of $A$ and shown to reproduce in a very satisfactory way the nucleon Spectral Function in $^3He$ and nuclear matter calculated within many-body approaches using realistic models of the $NN$ interaction.

\indent Let us now briefly describe the (non-relativistic) nucleon Spectral Function adopted in case of $^3He$ and $^3H$. For the former nucleus the ground component $P_0^N(p, E)$ is given by 
 \be
      P_0^p(p, E) & = & n_0^p(p) ~ \delta[E - E_{min}^{(^3He)}] ~ ,
     \nonumber \\
      P_0^n(p, E) & = & 0,
     \label{eq:P0_He3}
 \ee
where $n_0^p(p)$ is the proton momentum distribution corresponding to the $^3He$ to deuteron transition and $E_{min}^{(^3He)} \simeq 5.49 ~ MeV$. Note that the ground component $P_0^n(p, E)$ is identically vanishing because the residual $pp$ system does not possess any bound states. In case of $^3H$, since charge symmetry largely holds for the nuclear wave functions of $^3He$ and $^3H$, one can write: 
 \be
      P_0^n(p, E) & = & R(p) \cdot n_0^p(p) ~ \delta[E - E_{min}^{(^3H)}] ~ 
       , \nonumber \\
      P_0^p(p, E) & = & 0,
     \label{eq:P0_H3}
 \ee
where $n_0^p(p)$ is the same momentum distribution appearing in Eq. (\ref{eq:P0_He3}). The correction function $R(p)$, which includes the effects due to charge symmetry and charge independence breaking terms in the $NN$ (as well as $NNN$) interaction, turns out to be quite close to unity, namely within $2 \div 3 \%$ level of accuracy, as it can be estimated from the explicit calculations of $^3He$ and $^3H$ wave functions carried out in Ref. \cite{Nogga}. For $n_0^p(p)$ we use the simple parameterization obtained in Ref. \cite{2NC}(b) in case of the $RSC$ model \cite{RSC} of the $NN$ interaction. Note that $E_{min}^{(^3H)} \simeq 6.26 ~ MeV \neq E_{min}^{(^3He)}$ because of the different values of the experimental binding energies of $^3He$ and $^3H$.

\indent As for the correlated part $P_1^N(p, E)$ for $^3He$ we adopt the model of Ref. \cite{2NC} and the parameterizations of $n_1^{N=n, p}(p)$ corresponding to the $RSC$ interaction. For $^3H$ we follow the same logic of Eq. (\ref{eq:P0_H3}) and estimate the correction function $R(p)$ from the calculations of Ref. \cite{Nogga}. An additional difference in the correlated parts may arise from the different values of the threshold for the three-body break-up $E_{thr}$ (cf. \cite{2NC}), namely: $E_{thr} = 7.72$ and $8.48 ~ MeV$ in $^3He$ and $^3H$, respectively.

\indent In Fig. 1 we have reported the results for two ratios of light-cone momentum distribution functions, namely $f^p_{^3He} / f^n_{^3H}$ (solid line) and $f^n_{^3He} / f^p_{^3H}$ (dashed line), estimated according to the results of Ref. \cite{Nogga}. It can clearly be seen that the corrections to the charge independence in case of $^3He$ and $^3H$ nuclear wave functions are typically at the level of $2 \div 3 \%$.

\begin{figure}[htb]

\centerline{\epsfxsize=16cm \epsfig{file=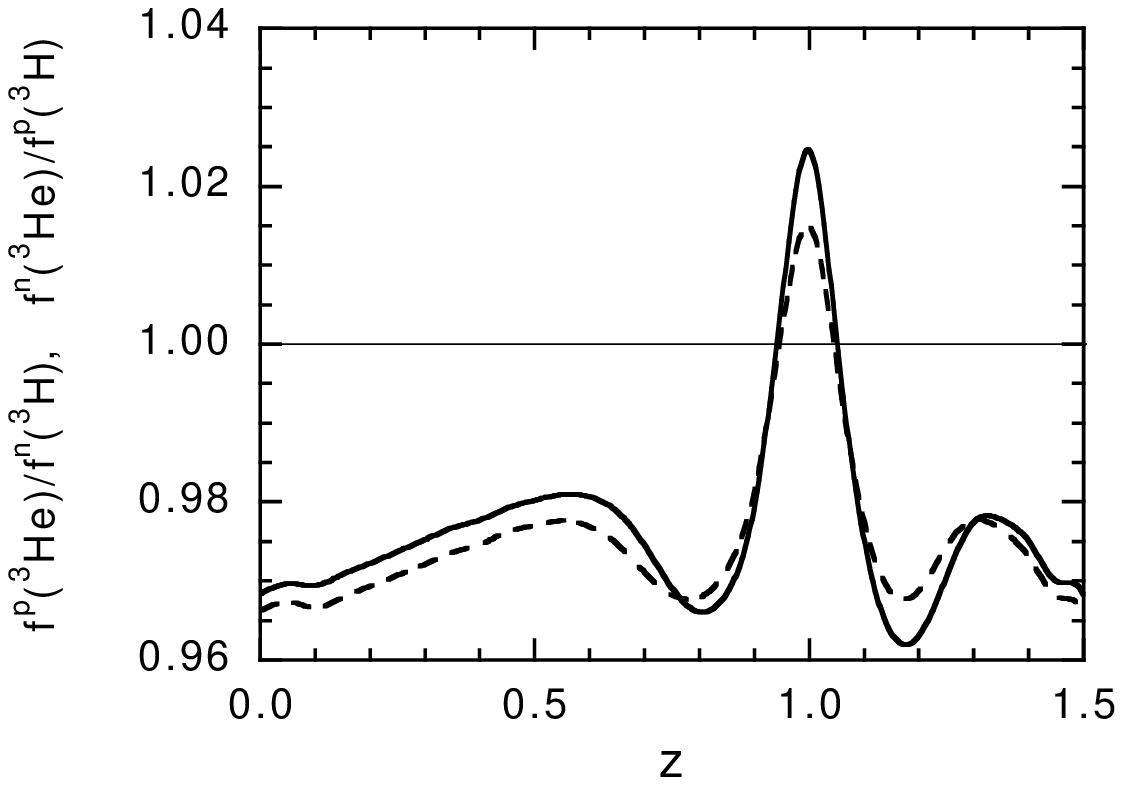}}

{\small {\bf Figure 1.} The ratio of the light-cone momentum distribution functions: $f^p_{^3He}/f^n_{^3H}$ (solid line) and $f^n_{^3He}/f^p_{^3H}$ (dashed line), estimated according to the results of Ref. \cite{Nogga}.}

\end{figure}

\indent Next we want to estimate the uncertainty introduced by the above-mentioned normalization procedures, given by Eqs. (\ref{fsnorm}) and (\ref{clnorm}). In Fig. 2 we have reported the results of our calculation of the proton and neutron light-cone momentum distributions in $^3He$ according to Eqs. (\ref{clnorm}) and (\ref{eq:fN}) (solid and dashed curves for proton and neutron, respectively) and according to Eqs. (\ref{nlcd}) and (\ref{fsnorm}) (solid and dashed curves with circles for proton and neutron, respectively). It turns out that the two different normalization prescriptions can substantially differ at small values of the light-cone fraction $z$ and this may represent a potential source of uncertainty.

\begin{figure}[htb]

\centerline{\epsfxsize=16cm \epsfig{file=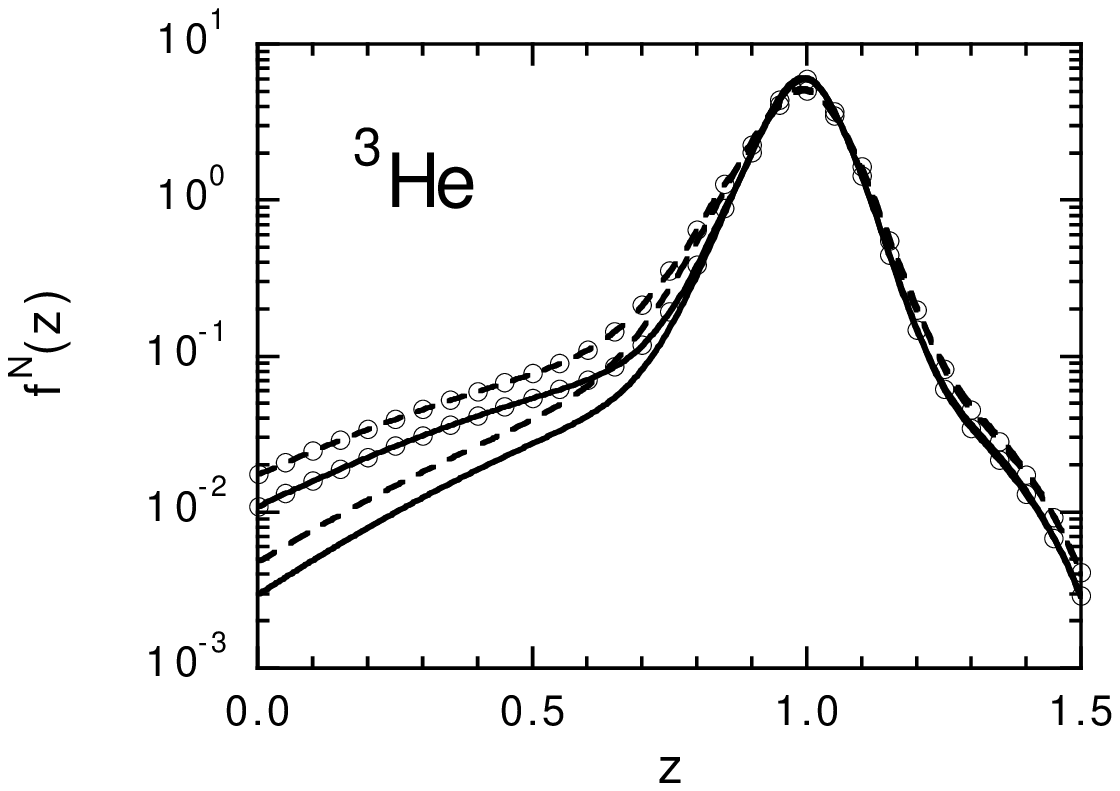}}

{\small {\bf Figure 2.} The nucleon light-cone momentum distribution in $^3He$, corresponding to the $RSC$ model \cite{RSC} of the $NN$ interaction, adopted in this work. The dashed and solid lines correspond to the neutron and proton momentum distribution, respectively. Lines with open circles correspond to the calculation with the normalization scheme of Eq. (\ref{fsnorm}), and those without open circles to the normalization scheme of Eq. (\ref{clnorm}).}

\end{figure}

\indent From Fig. 2 it can be seen also that within each normalization prescription $f^p(z) \neq f^n(z)$. Such a difference is driven by the presence in the three-nucleon wave function of a mixed-symmetry $S'$-wave component as well as of $P$- and $D$-waves arising from the spin-spin, spin-orbit and tensor terms of the $NN$ interaction, respectively. According to the results of sophisticated solutions of the three-nucleon ground states both with and without charge symmetry (and charge independence) breaking terms \cite{Nogga}, the probabilities of the $S'$, $P$ and $D$ partial waves are: $P_{S'} \simeq 1.2 \div 1.5 \%$, $P_P \lsim 0.2 \%$ and $P_D \simeq 7 \div 9 \%$, depending on the specific model adopted for the $NN$ (as well as $NNN$) interaction. Therefore, we do not expect that the dependence of $f^N(z)$ on the particular nuclear force model could be significant for the estimate of the deviation of the super-ratio (\ref{eq:superatio}) from unity, as it is also suggested by the results already obtained in Ref. \cite{Wally}. Finally, we want to stress that our results for the $EMC$ ratio (\ref{eq:REMC}) obtained within the convolution formula (\ref{eq:convolution}) for mirror $A = 3$ nuclei, have been positively checked against the corresponding results of Ref. \cite{CL} and \cite{Pace}, where the nucleon Spectral Function obtained from few-body variational techniques in case of the $RSC$ potential has been employed; we have found that the differences do not exceed $\simeq 1 \%$ in the whole $x$-range of interest in this work, viz. $0.3 \lsim x \lsim 0.9$.

\subsection{Light-Cone Approach}

\indent In this subsection we discuss the formalism of $LC$ dynamics in the approximation where the relevant degrees of freedom in the nuclear medium are the nucleons only, carrying therefore the total momentum of the nucleus. The structure function in the $LC$ framework can be represented as follows \cite{FS,FSS90}:
 \be
      F_2^A(x, Q^2) & = & \sum_{N = 1}^A \int {dz \over z^2} d^2p_{\perp} ~ 
      \rho^{LC}_N(z, p_{\perp}) ~ F_2^N(<\tilde{x}>, Q^2) {\nu \over 
      <\tilde{\nu}>} \nonumber \\ 
      & \cdot & \left[ ({M_A \over AM})^2 (1 + \mbox{cos} \delta)^2 (z + 
      \alpha_q {M <\nu'> \over Q^2})^2 + {p_{\perp}^2 \over 2 M^2} 
      \mbox{sin}^2\delta \right], 
      \label{F2LC} 
 \ee
where all the quantities in the r.h.s, except $\rho^{LC}_N(z, p_{\perp})$, are defined in subsection \ref{NSF}. In the Bjorken limit one obtains:
 \be
      F_2^A(x, Q^2) = \sum_{N = 1}^A \int_x^A {dz \over z}  f_N^{LC}(z) ~ 
      F_2^N({x \over z}, Q^2)
      \label{F2BLLC},
\ee
where $f_N^{LC}(z)$ is related to $\rho^{LC}_N(z, p_{\perp})$ according to Eq. (\ref{nlcd}). The quantity $\rho^{LC}_N(z, p_{\perp})$ represents the nucleon $LC$ density matrix in the nuclear medium. This function satisfies two sum rules; namely, from baryon charge conservation one has
 \be
       \int {dz \over z} d^2p_{\perp} ~ \rho^{LC}_N(z, p_{\perp}) = 1 ~, 
       \label{bchc}
 \ee
while the momentum sum rule requires
 \be
       \int {dz \over z} d^2p_{\perp} ~ z \rho^{LC}_N(z, p_{\perp})  = 1.
       \label{msm}
 \ee
Note that the last sum rule is not directly satisfied in the $VNC$ model, but it can be restored if mesonic degrees of freedom  are explicitly introduced.

\indent In general  $\rho^{LC}_N(z, p_{\perp})$ is not known for nuclei with $A \ge 3$. However, for numerical calculations one can proceed using the two following observations. First, in the non-relativistic limit (applicable for $0.7 \div 0.8 \lsim z \lsim 1.2 \div1.3$ and $p_\perp \lsim k_F \approx 200 \div 300 MeV/c$)  the density $\rho^{LC}_N(z, p_{\perp})$ can be related to the non-relativistic nucleon momentum distribution $n^N(p)$ as
  \be
      \rho^{LC}_N(z, p_{\perp}) \mid_{z \approx 1 - {p_z \over M}} \approx M
      ~ n^N(p)
      \label{nrl}.
 \ee
Second, within the two-nucleon correlation model \cite{2NC} one can relate the high momentum tail of the nuclear $LC$ density matrix to the two-nucleon density matrix on the light cone \cite{FS81}:
 \be 
      \rho^{LC}_N(z, p_{\perp}) \approx a_2(A) ~ \rho^{LC}_{NN}(z, 
      p_{\perp}) = {E_k ~ a_2(A) ~ n_{NN}(k) \over 2 - z}
       \label{2NN}
 \ee
where $E_k = \sqrt{M^2+k^2}$  and \cite{FS81} 
 \be
       k = \sqrt{{M^2 + p_{\perp}^2 \over z(2 - z)} - M^2}.
       \label{k}
 \ee
The observation \cite{2NC} that the two-nucleon correlations define the high momentum tail of the nuclear momentum distribution at momenta ($> k_F$) allows to replace $n^N(k)$ in Eq. (\ref{2NN}) by $a_2(A) ~ n_{NN}(k)$ at $k > k_F$, where $a_2(A)$ characterizes the probability to find a two-nucleon correlation in the high momentum tail of the nucleon momentum distribution in the nucleus $A$\footnote{An estimate of $a_2(A)$ for a variety of nuclei can be found in Ref. \cite{2NC}.}. Finally in the kinematical range where two-nucleon correlations dominate the proton-neutron pair, one can replace $n_{NN}(k)$ by the squared $LC$ wave function $|\Psi_D(k)|^2$ of the deuteron \cite{FS81}.

\indent The $LC$ many-nucleon approximation for the nuclear wave function leads to a prediction for the $F_2^A / F_2^N$ ratio which qualitatively contradicts the $EMC$ effect for $x \gsim 0.5$. This reflects the need to include explicitly non-nucleonic degrees of freedom in nuclei in order to explain the $EMC$ effect. In the $LC$ approximation a natural explanation is offered by the deformation of the quark wave function in the bound nucleon  which will be considered in the subsection \ref{csm}. In the following Section we will use the $LC$ model to illustrate the magnitude of the Fermi motion effects on the super-ratio (\ref{eq:superatio}).

\section{Numerical Estimates of Inelastic Cross Section}

\indent To check the reliability of our assumptions in the derivation of Eq. (\ref{F2IABC}) as well as of the models used for the three-nucleon Spectral Function, we first compare our calculations with the experimental data at moderate values of $Q^2 \sim 2 \div 3 ~ (GeV/c)^2$. At these values of $Q^2$ the nuclear modification of the valence quark distributions (the $EMC$ effect) is expected to be small and thus the comparison with the data will allow to check the validity of Eqs. (\ref{F2IABC}-\ref{rho}). Figure 3 presents the comparison of our calculations with the experimental data of Ref. \cite{Day}, where for the nucleon structure function $F_2^N(x, Q^2)$ we have used the parameterization of Ref. \cite{Bodek} containing also the contribution of nucleon resonances.  For the evaluation of $F_1^A$ [see Eq. (\ref{F1IA})] we have used the relation $F_1^N = F_2^N [1 + (2Mx / \nu)] / 2x (1 + R)$ with $R = 0.18$. The comparison clearly demonstrates that Eq. (\ref{F2IABC}) is a good starting point for the discussion of higher $Q^2$ regime.

\begin{figure}[htb]

\centerline{\epsfxsize=16cm \epsfig{file=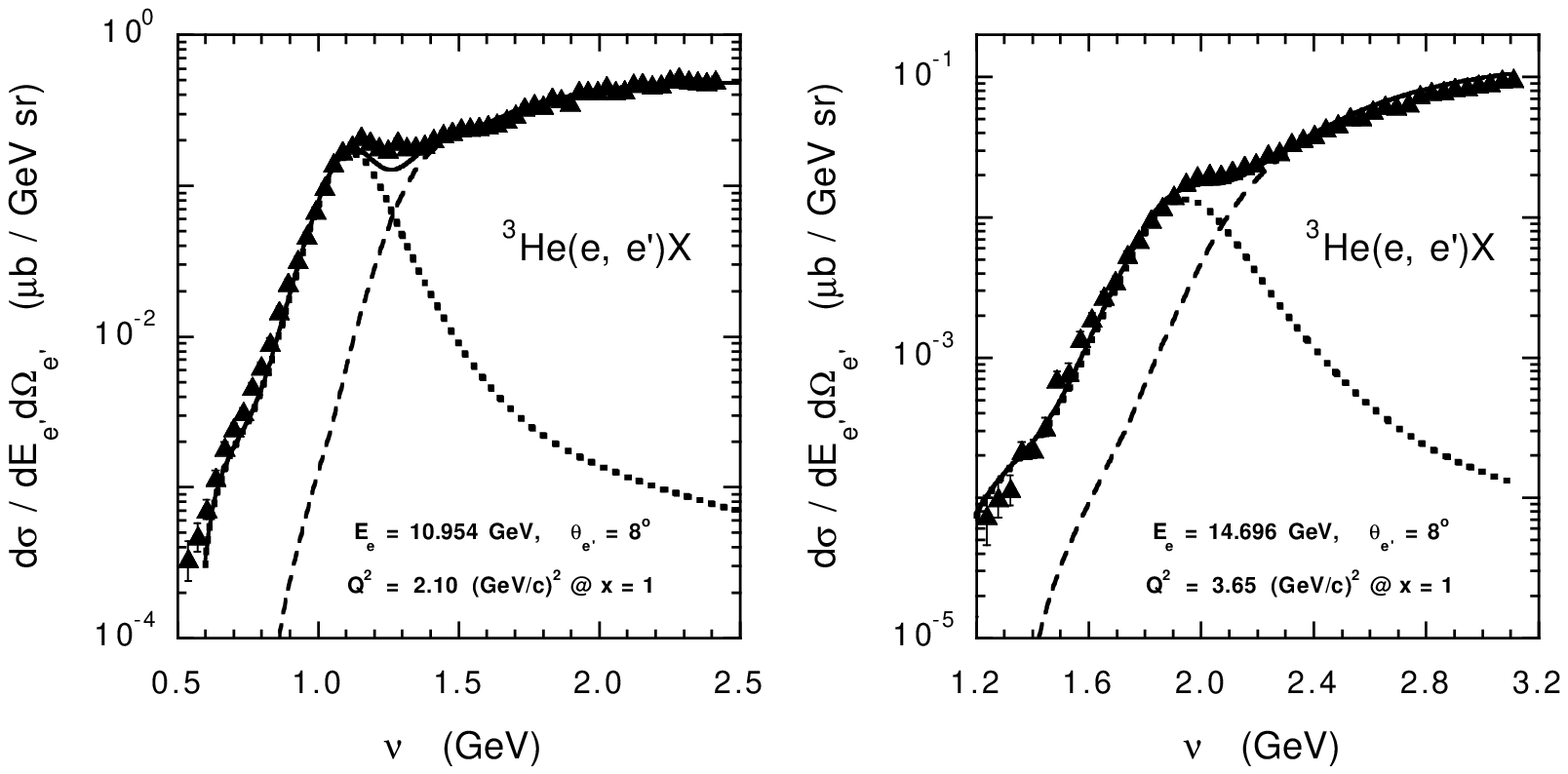}}

{\small {\bf Figure 3.} The cross section of inclusive $^3He(e,e')X$ scattering as a function of energy transfer $\nu$. Dashed line - inelastic contribution calculated according to Eq. (\ref{F2IABC}) and adopting for the nucleon structure function the parameterization of Ref. \cite{Bodek}, which includes nucleon resonances. Dotted line - quasi-elastic contribution calculated according to Ref. \cite{QE}. Solid line - total cross section. The experimental data (full triangles) are from Ref. \cite{Day} and the kinematical conditions are shown in the insets.}

\end{figure}

\indent At larger $Q^2$ the first question we want to address is how fast the Bjorken limit is established and how much the nuclear recoil effects accounted for in Eq. (\ref{F2IABC}) are important. The finite $Q^2$ effects are governed by the scale of the target-mass corrections ($\sim M^2 / Q^2$ as well as the factors proportional to $Q^2 / \nu^2$ and $p_{\perp}^2 / Q^2$). In Fig. 4 we compare the results obtained for both the $EMC$ ratio and the super-ratio for $^3He$ and $^3H$ targets at $Q^2 = 10 ~ (GeV/c)^2$, calculated within Eq. (\ref{F2IABC}) and the convolution formula (\ref{eq:convolution}). It can clearly be seen that, while for the $EMC$ ratio (\ref{eq:REMC}) the convolution formula works within a $2 \div 3 \%$ level of accuracy at large $x$, in case of the super-ratio (\ref{eq:superatio}) the differences between the non-Bjorken and the Bjorken limits cancel out almost completely.

\begin{figure}[htb]

\centerline{\epsfxsize=16cm \epsfig{file=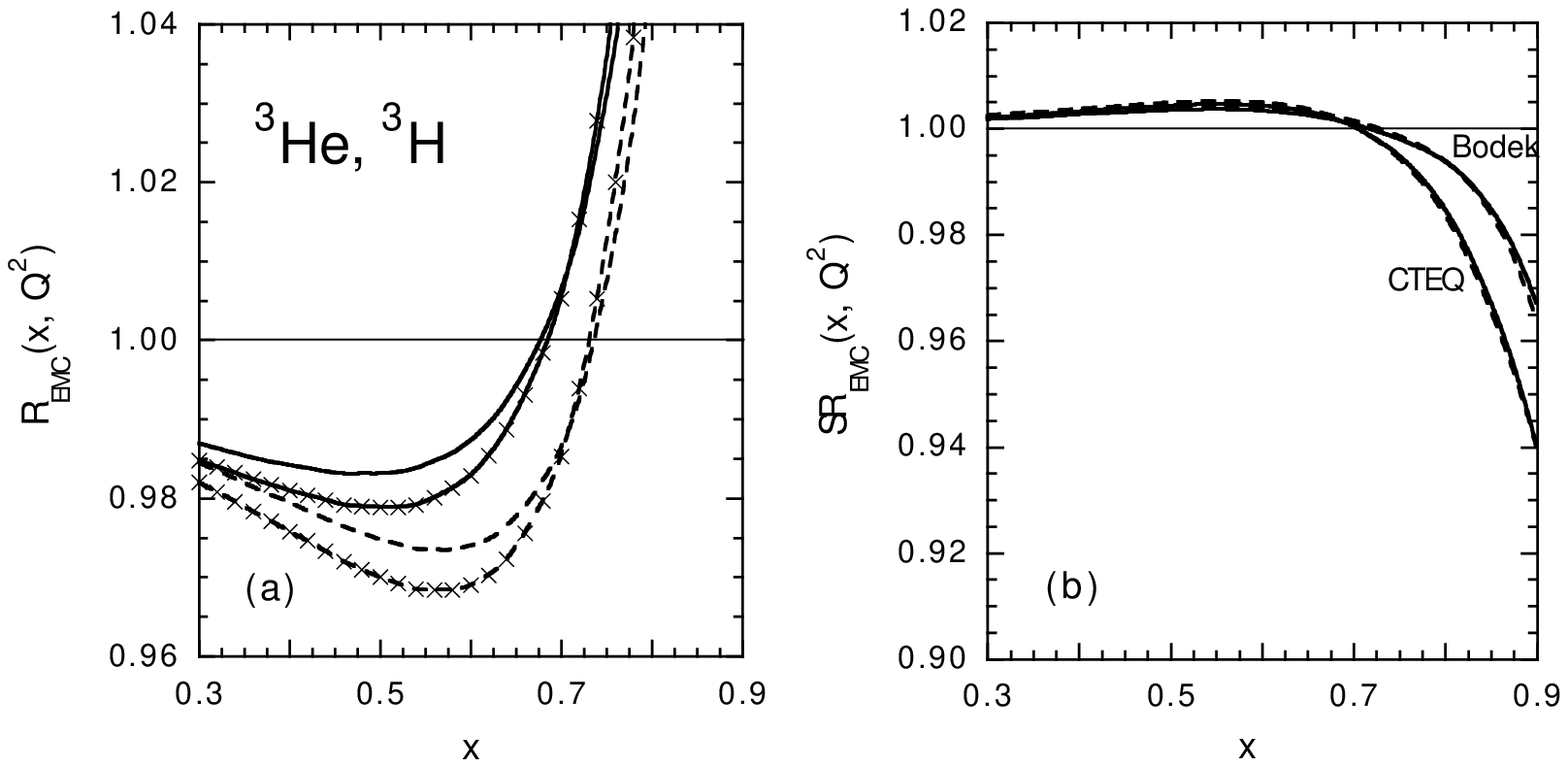}}

{\small {\bf Figure 4.} The $x$ dependence of  ${\cal{R}}_{EMC}^A$ (a) and ${\cal{SR}}_{EMC}$ (b) for $^3He$ and $^3H$ targets at $Q^2 = 10 ~ (GeV/c)^2$. Dashed and solid lines are the results obtained using Eq. (\ref{F2IABC}) calculated without invoking the Bjorken limit, and Eq. (\ref{eq:convolution}) using the Bjorken limit, respectively. Lines marked by crosses correspond to $^3H$ target, unmarked lines to $^3He$ target. In (a) and (b) the $CTEQ$ set of $PDF$'s from Ref. \cite{CTEQ} is adopted, while in (b) the results obtained using the parameterization of Ref. \cite{Bodek} for the nucleon structure function $F_2^N(x, Q^2)$ are also reported. The charge-symmetry breaking effects shown in Fig. 1 are included in the calculations.}

\end{figure}

\indent The next question is the expected uncertainty on the $EMC$ ratio due to the different normalization procedures of the nucleon Spectral Function discussed in the previous Section. In Fig. 5 we compare the calculation of ${\cal{R}}_{EMC}^{A = 3}$ and ${\cal{SR}}_{EMC}$ performed within the convolution approximation using the two different schemes of normalization given by Eqs. (\ref{fsnorm}) and (\ref{clnorm}). It can be seen that ${\cal{R}}_{EMC}^{A = 3}$ exhibits some sensitivity to the choice of the normalization scheme, while the differences in the calculated ${\cal{SR}}_{EMC}$ are well below $\sim 1 \%$.

\begin{figure}[htb]

\centerline{\epsfxsize=16cm \epsfig{file=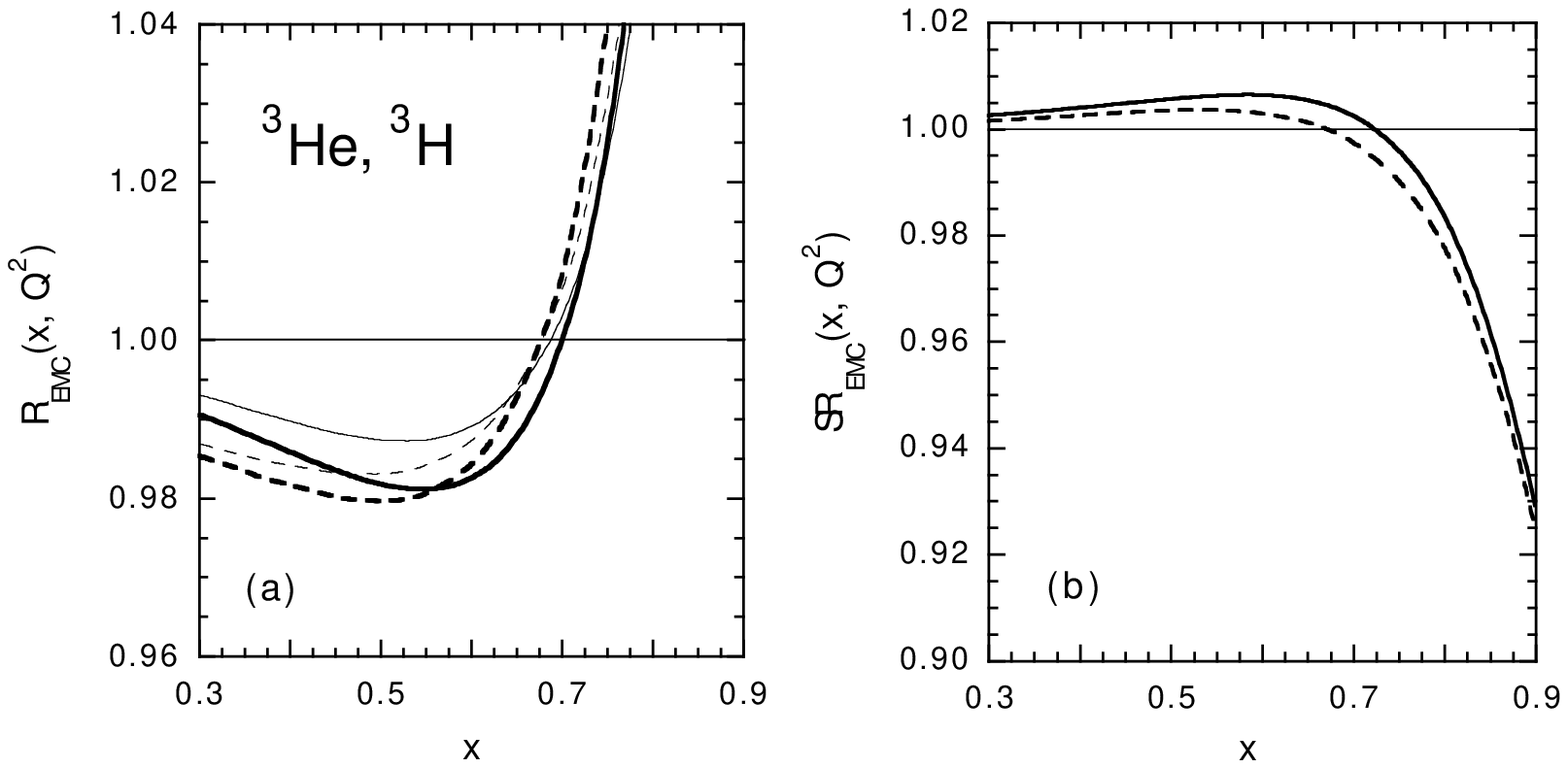}}

{\small {\bf Figure 5.} (a) The $EMC$ ratio [Eq. (\ref{eq:REMC})] in $^3He$ (thin lines) and in $^3H$ (thick lines) vs. $x$ at $Q^2 = 10 ~ (GeV/c)^2$. (b) The super-ratio [Eq. (\ref{eq:superatio})] vs. $x$. Dashed and solid lines correspond to the results obtained using the normalization schemes of Eqs. (\ref{fsnorm}) and (\ref{clnorm}), respectively. The $CTEQ$ set of $PDF$'s from Ref. \cite{CTEQ} has been adopted as input for the nucleon structure function $F_2^N(x, Q^2)$. The charge-symmetry breaking effects shown in Fig. 1 are not included in the calculations.}

\end{figure}

\indent Next we  address the sensitivity of the super-ratio (\ref{eq:superatio}) to the particular choice of the $PDF$ parameterization in the nucleon. To this end we have calculated the super-ratio (\ref{eq:superatio}) via the convolution formula (\ref{eq:convolution}) using the proton and neutron $LC$ momentum distributions (\ref{eq:fN}) with the normalization procedure given by Eq. (\ref{clnorm}). The $RSC$ model \cite{RSC} of the $NN$ interaction adopted in the calculation yields $C_n = 1.048$ and $C_p = 1.033$ for the normalization constants in  Eq. (\ref{clnorm}). We have neglected the charge-symmetry breaking effects shown in Fig. 1 and  we have used different parameterizations of the nucleon structure function $F_2^N(x, Q^2)$ taken at $Q^2 = 10 ~ (GeV/c)^2$, namely the $GRV$ set \cite{GRV} of $PDF$'s and the $SLAC$ fit of Ref. \cite{SLAC}. Both the $GRV$ and $SLAC$ parameterizations are constructed in a such a way that the neutron to proton structure function ratio reaches the "non-perturbative prediction" $1 / 4$ as $x \to 1$ (see Fig. 6(a)). We have therefore applied to the $GRV$ and $SLAC$ structure functions an {\em ad hoc} modification in the form of a distortion of the $d$-quark distribution limited only at large $x$ (i.e., $x \gsim 0.7$), viz.: $d(x) \to d(x) + 0.1 x^4 (1 + x) u(x)$. Such a modification has been directly implemented in the $GRV$ set of $PDF$'s, while in case of the $SLAC$ parameterization we have considered the following replacements: $F_2^p(x, Q^2) \to F_2^p(x, Q^2) \{ 1 + 0.1 x^4 (1 + x) / 4 \}$ and $F_2^n(x, Q^2) \to F_2^n(x, Q^2) \{ 1 + 4 \cdot  0.1 x^4 (1 + x)  \}$. In both cases the $n/p$ ratio of the modified structure functions goes to the "$pQCD$ prediction" $3 / 7 \simeq 0.43$ as $x \to 1$ (see Fig. 6(a)). Since the proton structure function is dominated at large $x$ by the $u$-quark distribution, the above-mentioned modification does not change significantly the shape of the proton structure function; the effects are larger on the neutron structure function, but by construction they are limited in the region $x \gsim 0.7$ (see Fig. 6(a)).

\begin{figure}[htb]

\centerline{\epsfxsize=16cm \epsfig{file=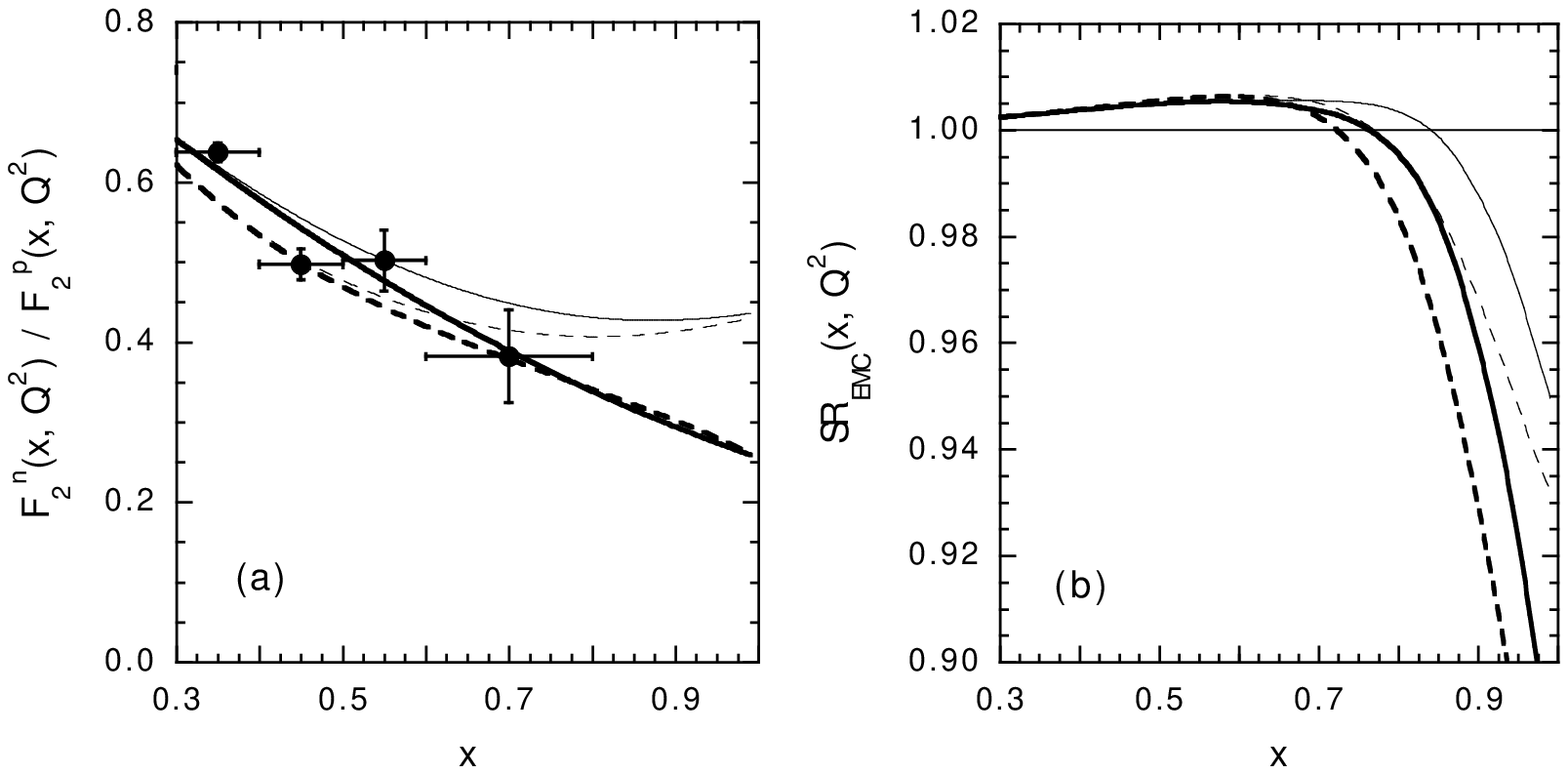}}

{\small {\bf Figure 6.} (a) Neutron to proton structure function ratio, $F_2^n(x, Q^2) / F_2^p(x, Q^2)$, vs. the Bjorken variable $x$ at $Q^2 = 10 ~ (GeV/c)^2$. The full dots are the $NMC$ data points as given in Ref. \cite{Arneodo}. Thick dashed and solid lines correspond to the $GRV$ set \cite{GRV} of $PDF$'s and to the $SLAC$ parameterization of Ref. \cite{SLAC}. Thin lines are the modified $GRV$ and $SLAC$ fits, as described in the text. (b) Super-ratio [Eq. (\ref{eq:superatio})] of the $EMC$ effects in $A = 3$ mirror nuclei. The meaning of the lines is the same as in (a). Using the $CTEQ$ parameterization \cite{CTEQ} one obtains results  very similar to those reported for the $GRV$ set \cite{GRV} of $PDF$'s.}

\end{figure}

\indent The results obtained for the super-ratio (\ref{eq:superatio}) are shown in Fig. 6(b). It can be clearly seen that the deviation of the super-ratio from unity is small (less than $1 \%$) up to $x \simeq 0.75$, while it increases rapidly as $x \gsim 0.75$ and depends strongly on the large-$x$ behavior of the $n/p$ ratio.  Our conclusion is that the $VNC$ model predicts a deviation of the super-ratio from unity within $1 \%$ only for $x \lsim 0.75$ in overall agreement with the results of Refs. \cite{Wally,Pace}. Note however that the $x$-shape and the average value of our results for the super-ratio are closer to the findings of Ref. \cite{Pace} (where a Spectral Function similar to the one of the present work is adopted) and differs from the results of Ref. \cite{Wally}, where larger deviations (up to $2 \%$) from unity were found. It is likely that the difference is related to the different Spectral Functions used in the present work and in Ref. \cite{Wally}, since the latter uses the $VNC$ model with  the same normalization scheme [Eq. (\ref{fsnorm}))] for the nucleon $LC$ momentum distribution.

\begin{figure}[htb]

\centerline{\epsfxsize=12cm \epsfig{file=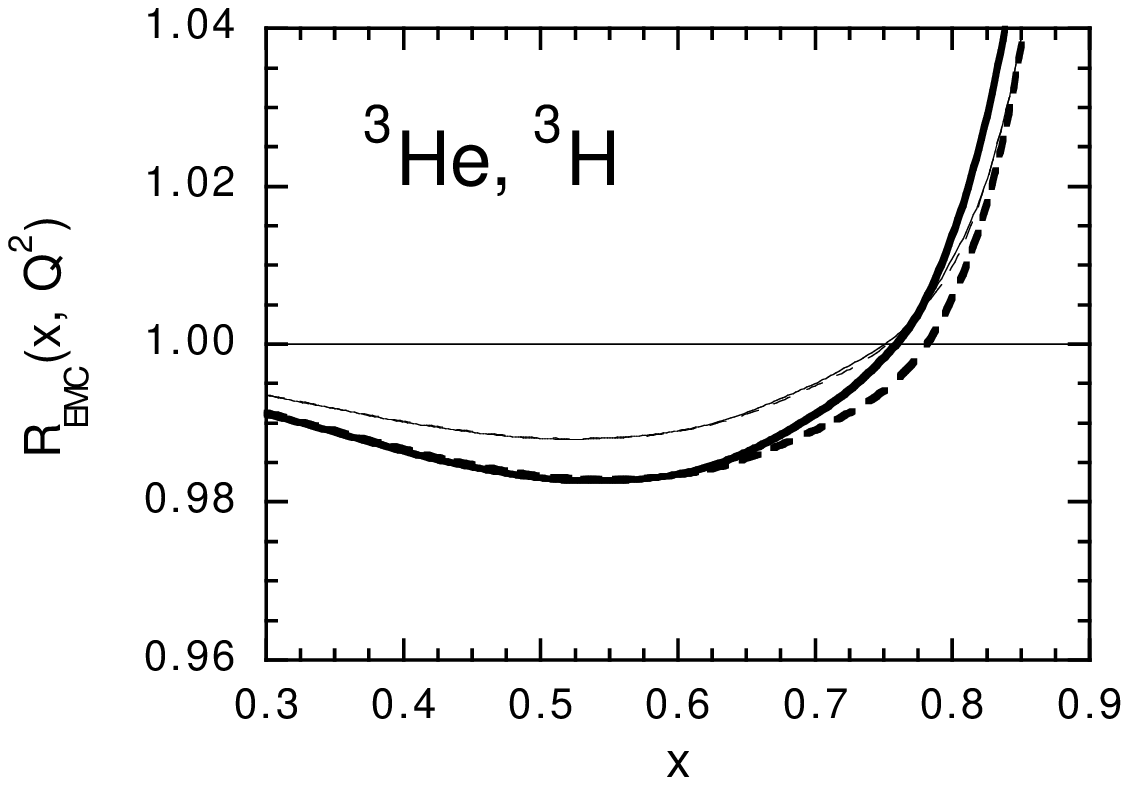}}

{\small {\bf Figure 7.} The $EMC$ ratio [Eq. (\ref{eq:REMC})] in $^3He$ (thin lines) and in $^3H$ (thick lines) as predicted by the convolution formula (\ref{eq:convolution}) at $Q^2 = 10 ~ (GeV/c)^2$. The solid lines correspond to the $SLAC$ \cite{SLAC} parameterization of the nucleon structure function, while the dashed lines are the results obtained using the modified $SLAC$ fit at large $x$ as described in the text. The thin dashed and solid lines are almost indistinguishable.}

\end{figure}

\indent The $x$-shape of the super-ratio ${\cal{SR}}_{EMC}(x, Q^2)$ shown in Fig. 6(b), can be better understood by looking at Fig. 7, where the $EMC$ ratio ${\cal{R}}_{EMC}(x, Q^2)$ in the two mirror $A = 3$ nuclei is separately reported. It can be seen that the convolution approach predicts a larger deviation from unity in $^3H$ than in $^3He$ for $x \lsim 0.75$. This is a direct consequence of the higher kinetic energy of the neutron (proton) with respect to the proton (neutron) in $^3He$ ($^3H$) due to the spin-flavor dependence of the nuclear force [see the discussion after Eq. (\ref{eq:mirror})]. The super-ratio ${\cal{SR}}_{EMC}(x, Q^2)$ is therefore larger than one up to $x \simeq 0.75$ (see Fig. 6(b)), but such a deviation from unity is small because the $EMC$ ratio itself is predicted to be quite small in the two mirror nuclei (less than a $1 \%$ effect) within the VNCM. For $x \gsim 0.75$ the $EMC$ ratio increases above unity very sharply; generally speaking, this is related to the fact that the nucleon structure function goes to zero as $x \to 1$, while the nuclear one is non-vanishing because of the Fermi motion of the nucleons in the nucleus. Moreover, the slope of the rise of ${\cal{R}}_{EMC}(x, Q^2)$ is larger in $^3H$ than in $^3He$ due to the  decrease of $F_2^n/F_2^p$ at $x \to 1$. Note also that the $EMC$ ratio in $^3H$ is sensitive to the modification of the $d /u$ ratio at large $x$, whereas the $EMC$ ratio in $^3He$ is not (see Fig. 7). Thus, for $x \gsim 0.75$ the super-ratio ${\cal{SR}}_{EMC}(x, Q^2)$ drops below one and becomes a rapidly varying function of $x$ with a remarkable sensitivity to the large-$x$ shape of the nucleon structure function.

\indent Finally the last source of uncertainty we want to consider within the convolution approximations is the difference between the predictions of the $VNC$ model and the $LC$ formalism. In Fig. 8 the predictions for $EMC$ ratio calculated within the $VNC$ model according to Eq. (\ref{F2IABC}) and the normalization scheme of Eq. (\ref{fsnorm}) are compared with the corresponding ones of the $LC$ approach [see Eq. (\ref{F2LC})]. In both cases we have adopted the $F_2^N$ parameterization of Ref. \cite{Bodek} , which contains the contribution of nucleon resonances. As it follows from Fig. 8(a) the $LC$ approximation predicts larger value of the $EMC$ ratio as compared with the $VNC$ model. As a result the super-ratio within the $LC$ approximation is smaller (closer to one) as compared with the prediction of the $VNC$ model. Note also that the effects of nucleon resonances are still visible in Fig. 8(a) for $x \gsim 0.8$, corresponding to $W < 2 ~ GeV$ at $Q^2 = 10 ~ (GeV/c)^2$. Therefore, if one wants to investigate only the leading twist of the nucleon structure function, one can either limit the range of values of $x$ or increase sufficiently the value of $Q^2$.

\indent To sum up this Section, we conclude that all the considered uncertainties within the convolution approximation, in which no nuclear modification of bound nucleons is considered, do not yield deviations of the super-ratio (\ref{eq:superatio}) from unity larger than $1 \%$ at $x \lsim 0.75$ ($2 \%$ at $x \lsim 0.8$).

\begin{figure}[htb]

\centerline{\epsfxsize=16cm \epsfig{file=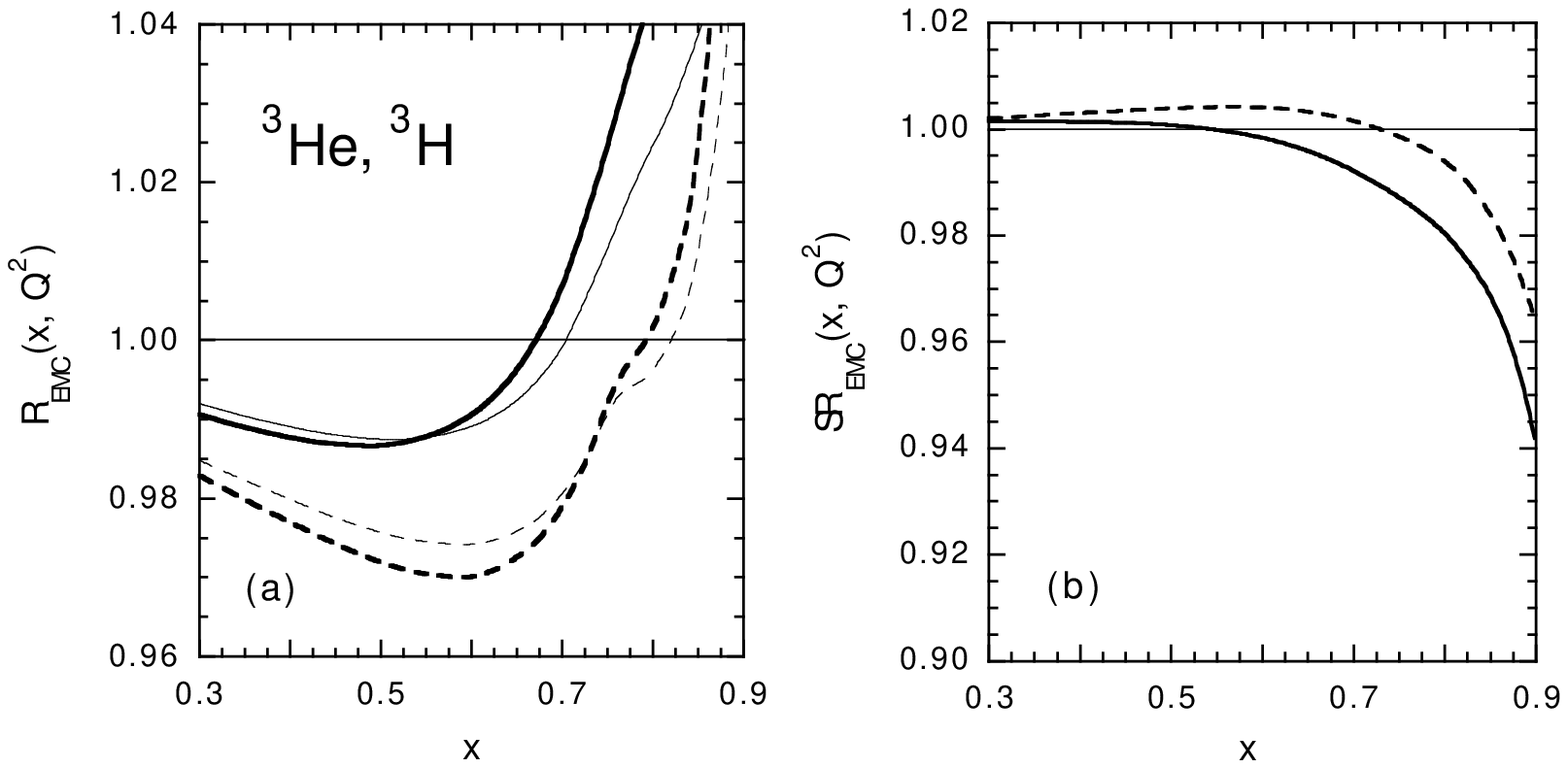}}

{\small {\bf Figure 8.} (a) The $EMC$ ratio [Eq. (\ref{eq:REMC})] in $^3He$ (thin lines) and in $^3H$ (thick lines) vs. $x$ at $Q^2 = 10 ~ (GeV/c)^2$. (b) The super-ratio [Eq. (\ref{eq:superatio})] vs. $x$. Dashed and solid lines correspond to the results obtained using the $VNC$ model and the $LC$ formalism, respectively. The parameterization of Ref. \cite{Bodek}, which includes nucleon resonances, has been adopted as input for the nucleon structure function $F_2^N(x, Q^2)$. The charge-symmetry breaking effects shown in Fig. 1 are not included in the calculations.}

\end{figure}

\section{Models of the $EMC$ Effect with Modifications of the Bound Nucleon Structure Function}

\indent Although within the $VNC$ model and the $LC$ approach the nuclear corrections to the super-ratio (\ref{eq:superatio}) are $\approx 1\%$  at $x \lsim 0.75$, it is hardly safe to treat this as an ultimate estimate of the nuclear effects. The $VNC$ model is just one of the many models of the $EMC$ effect. Also, literally, the $VNC$ model predicts the parton densities to violate the momentum sum rule (for instance by $\sim 5\%$ for an iron target). When this feature is fixed by adding mesonic (pionic) degrees of freedom, one predicts an enhancement of the antiquark distributions in nuclei at $x \ge 0.05$ which grossly contradicts the Drell-Yan data \cite{DY}. It is also well known that the convolution approximations underestimate significantly the $EMC$  effect at large $x$  (cf., e.g., Refs. \cite{FS,CL} and \cite{Gomez}). Experimental data are available for a variety of nuclei and in Fig. 9 we have limited ourselves to the cases of $^4He$ and $^{56}Fe$. The convolution formula within the $VNC$ model (\ref{eq:convolution}) has been evaluated adopting for the nucleon Spectral Function $P^N(k, E)$ the model of Ref. \cite{2NC} and our results turn out to be in agreement with the findings of Ref. \cite{CL}. From Fig. 9 it can clearly be seen that the convolution approach is not able to reproduce the minimum of the $EMC$ ratio around $x \approx 0.7$ as well as the subsequent sharp rise at larger $x$. Note that the disagreement is even larger within the $LC$ approximation (see dashed curves in Fig. 16).

\begin{figure}[htb]

\centerline{\epsfxsize=16cm \epsfig{file=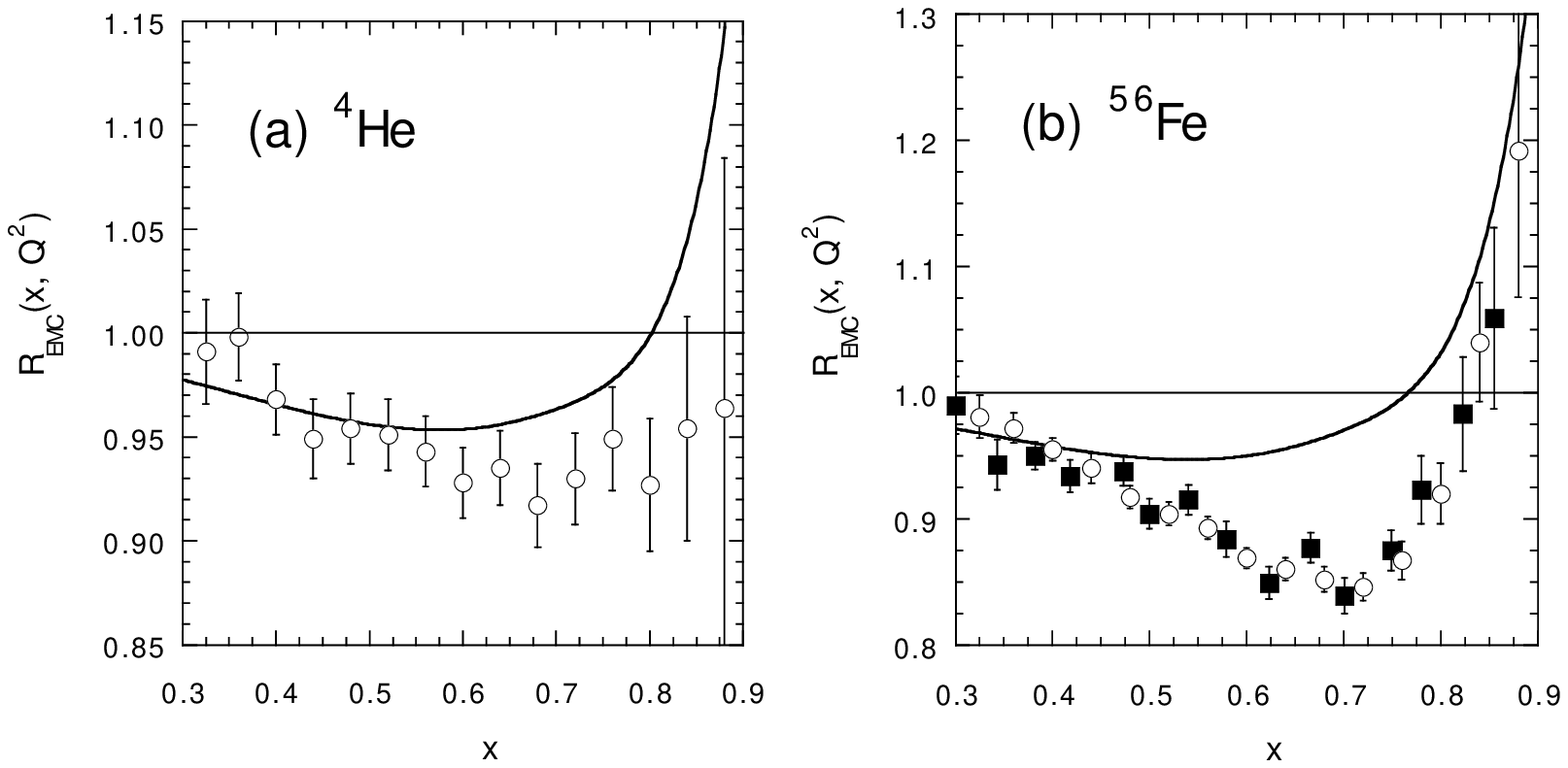}}

{\small {\bf Figure 9.} The $EMC$ ratio [Eq. (\ref{eq:REMC})] in $^4He$ (a) and $^{56}Fe$ (b) at $Q^2 = 10 ~ (GeV/c)^2$. Open dots are data from Ref. \cite{Gomez}, while in (b) the full squares are from Ref. \cite{Arnold}. The solid lines are the results of the convolution formula (\ref{eq:convolution}), calculated adopting the $SLAC$ \cite{SLAC} parameterization of the nucleon structure function $F_2^N(x, Q^2)$ and the model of Ref. \cite{2NC} for the nucleon Spectral Function $P^N(k, E)$.}

\end{figure}

Therefore, it is reasonable to expect that the results of the convolution approximation for mirror $A = 3$ nuclei can suffer the same drawback. Moreover it is very important to asses any isospin dependence of EMC effect in order to extract in a reliable way the neutron structure function from $^3He$ and $^3H$ data. An isospin dependence for EMC effect is naturally expected from the differences in the relative motion of $pn$ and $nn$ ($pp$) pairs in $^3H$ ($^3He$). The results obtained for $^3H$ in Ref. \cite{Pieper} in case of the Argonne $V18$ + Urbana $IX$ models of the $NN$ and $NNN$  interactions, are reported in Fig. 10. It can clearly be seen that, since the interaction of a $pn$ pair is more attractive than the one of a $nn$ pair, the proton is closer to the $^3H$ center-of-mass than the neutron. The corresponding root mean square radius turns out to be: $\sqrt{<r_{pn}^2>} \simeq 2.5 ~ fm$  and $\sqrt{<r_{nn}^2>} \simeq 2.8 ~ fm$. As a consequence, the overlapping probability is larger for a $pn$ pair than for a $nn$ pair. As a matter of fact, from Fig. 10 it follows that the partially integrated probability to find a $NN$ pair with $r_{NN} \leq 1 ~ fm$, is $\sim 40\%$ larger for a $pn$ pair than for a $nn$ pair. We stress that this is a very important isospin effect in mirror $A = 3$ nuclei.

\begin{figure}[htb]

\centerline{\epsfxsize=16cm \epsfig{file=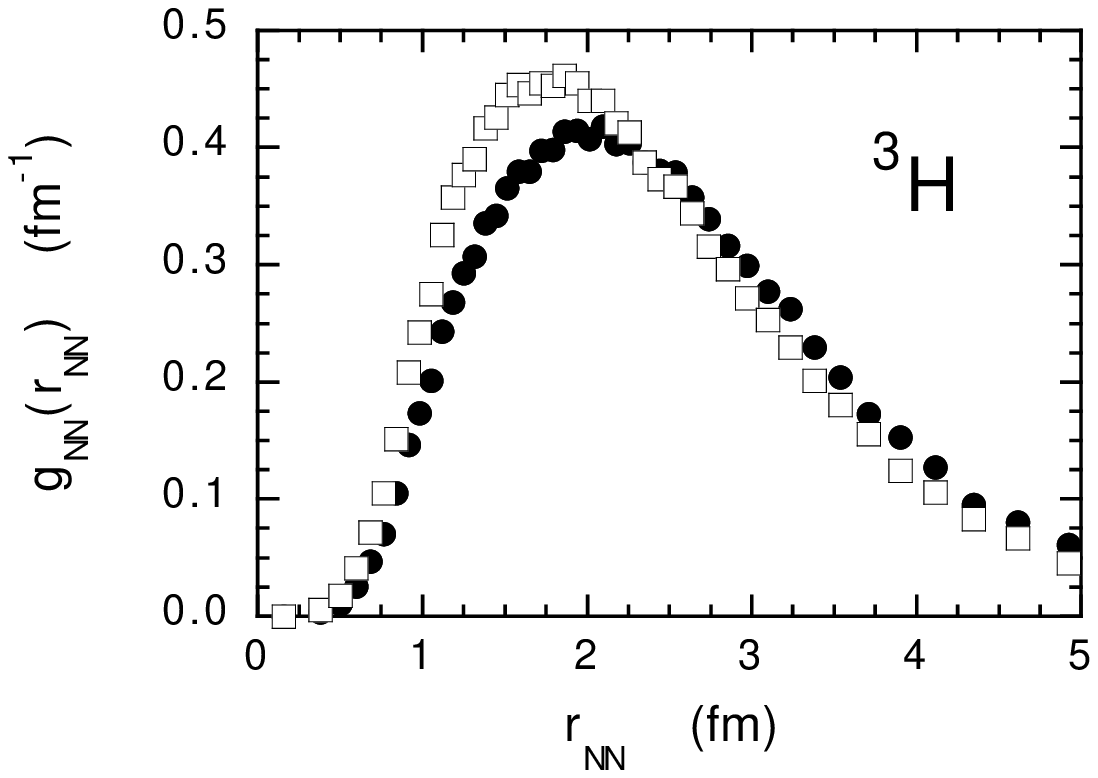}}

{\small {\bf Figure 10.} The distributions $g_{NN}(r_{NN})$ of the relative motion of a $nn$ pair (full dots) and of a $pn$ pair (open squares) in $^3H$, as a function of the relative distance $r_{NN}$ between the $NN$ pair. The results reported correspond to the the Argonne $V18$ and Urbana $IX$ models of the $NN$ and $NNN$ interactions, respectively, obtained using the Green Function Monte Carlo method of Ref. \protect\cite{Pieper}. The distributions are normalized as: $\int_0^{\infty} dr_{NN} g_{NN}(r_{NN}) = 1$.}

\end{figure}

{\em \indent Thus in order to draw final conclusion about the size of the deviation of the super-ratio ${\cal{SR}}_{EMC}(x, Q^2)$ from unity we should investigate effects beyond those predicted by the convolution approach.}

\indent The discovery of the  $EMC$ effect at large $x$ has triggered a huge theoretical effort which has led to the development of a large number of models (see, e.g., Refs. \cite{Close,Jaffe,FS85,Gomez} and references therein). In this work we will limit ourselves to consider some of these models, which are of interest for an estimate of the possible deviation of the super-ratio ${\cal{SR}}_{EMC}(x, Q^2)$ from unity. We will use the experimental points presented in Fig. 9 in order to constrain as much as possible the parameters of these models. Note that the experimental uncertainties on the $EMC$ ratio in $^4He$ are significantly larger than the corresponding ones in $^{56}Fe$; therefore, new measurements on $^4He$ targets with reduced errors will certainly help in improving our knowledge of the $EMC$ effect in light nuclei.

\subsection{Nuclear Density Model}

\indent It was argued in \cite{FS85,FS} that due to a diluteness of the nuclear systems the nuclear effects for the deviation of the nuclear structure function from the sum of the nucleon structure functions can be treated as a series in the powers of $k^2 / M^2$ and $\epsilon_A / M$. This approximation holds in a number of dynamical models, like the rescaling model \cite{Jaffe,Close,Pirner}, the six quark ($6q$) cluster model \cite{Vary,Carlson}, the color screening model (suppression of small size configurations in bound nucleons) \cite{FS85,FS,FJM} and pion models \cite{Berger}. Hence in the region of small enough $x$ (i.e, $x \lsim 0.7$), where terms $\propto  k^4 / M^4$ can be neglected, an approximate factorization should take place
 \be
       \overline{\cal{R}}_{EMC}^A(x,Q^2) - 1 = \beta(x, Q^2) ~ f(A),
       \label{factor}
 \ee
where $\overline{\cal{R}}_{EMC}^A(x, Q^2) \equiv F_2^A(x, Q^2) / [Z F_2^p(x, Q^2) +  N F_2^n(x, Q^2)]$, and 
 \be
      f(A) \propto <k^2> / M^2
      \label{k2}
 \ee
or to the average virtuality of the nucleon. Eq. (\ref{factor}) is in a very good agreement with the $SLAC$ data on the $A$-dependence of the $EMC$ effect. Numerical estimates using Eqs. (\ref{factor},\ref{k2}) and realistic deuteron and iron wave functions lead to \cite{FS85} 
\be
      {F_2^D(x, Q^2) \over F_2^p(x, Q^2) + F_2^n(x, Q^2)} - 1 \approx {1 
      \over 4} {F_2^{Fe}(x, Q^2) \over F_2^D(x, Q^2)} - 1.
\ee

\indent For $A \gsim 12$ one has approximately $<k^2> / M^2 \propto <\rho_A(r)> \equiv \rho(A)$, where $\rho(A)$ is the average nuclear matter density, leading to 
\be
       {\cal{R}}_{EMC}^A(x,Q^2) \equiv { \overline{\cal{R}}_{EMC}^A(x, Q^2) 
       \over  \overline{\cal{R}}_{EMC}^D(x, Q^2)} = \alpha(x, Q^2) ~ [1 + 
       \rho(A) \beta(x, Q^2)]~.
       \label{eq:density}
 \ee
Note in passing that such an approximation is definitely not applicable at very large $x$, since short-range correlations dominate for $x \gsim 1$ and therefore the relation ${\cal{R}}_{EMC}^A(x) \propto \rho_A$ is expected not to hold any more. Also one hardly can directly use this approximation for the deuteron since the notion of average nuclear density is not well defined in this case.

\indent Analysis of the data on the $EMC$ effect using Eq. (\ref{eq:density}) (including deuteron and $^4He$ data) has been carried out in terms of the average nuclear density $\rho(A)$ in Ref. \cite{Gomez}. The quantities $\alpha(x, Q^2)$ and $\beta(x, Q^2)$ were fitted  to the data; their values for various $x$-bins can be easily read off from Table IX of \cite{Gomez}. The nuclear density $\rho(A)$ was assumed to be given by : $\rho(A) = 3 A / 4 \pi R_e^3$ where $R_e = \sqrt{5/3} \cdot r_{c,A}$, with $r_{c,A}$ representing the r.m.s. electron scattering  (charge) radius of the nucleus. In particular, the values $\rho(^4He) = 0.089 ~ fm^{-3}$ and $\rho(^{56}Fe) = 0.117 ~ fm^{-3}$ were adopted in \cite{Gomez}. We will refer hereafter to Eq. (\ref{eq:density}) as the density model.

Note that the fit in Eq. (\ref{eq:density}) has been done in Ref. \cite{Gomez} using charge rather than matter radii of nuclei, which is a good approximation for large $A$ and $Z=N$ nuclei since in this case 
 \be
        <r^2_{c, A}> ~ = ~ <r^2_{matter,A}> + <r^2_{c, proton}> + <r^2_{c, 
        neutron}>
        \label{eq:radiiA}
 \ee
and 
 \be
        <r^2_{matter, A}> ~ \gg ~  <r^2_{c, proton}> ~, ~ <r^2_{c, neutron}>
        ~.
        \label{eq:largeA}
 \ee
For light isosinglet nuclei Eq. (\ref{eq:radiiA}) is expected to hold. However the predictions of the density model for light isosinglet nuclei should have a rather qualitative character since Eq. (\ref{eq:largeA}) does not hold for deuteron and barely holds for $^4He$ nucleus. Moreover, the step leading from Eqs. (\ref{factor},\ref{k2}) to Eq. (\ref{eq:density}) is not justified.

\indent We have mentioned above that the density model was proposed in Ref. \cite{FS85} only in case of sufficiently heavy nuclei. If we want to apply Eq. (\ref{eq:density}) to mirror $A = 3$ nuclei, the first question is which density we have to use. As already observed in Fig. 10, the neutron (proton) is closer to the $^3He$ ($^3H$) center-of-mass than the proton (neutron). This means that the neutron (proton) has more kinetic energy of the proton (neutron) in $^3He$ ($^3H$)\footnote{In what follows we neglect the small isospin violation driven by charge symmetry breaking effects in the relative motion of the $pp$ pair in $^3He$ and of the $nn$ pair in $^3H$.}. According to the $RSC$ interaction, the neutron in $^3He$ possesses on average about $25 \%$ kinetic energy more than the proton. Since the deviation of the $EMC$ ratio from unity may be related to the mean kinetic energy of the nucleon and to the derivatives of the nucleon structure function (cf. \cite{FS}), we expect a different $EMC$ effect in mirror $A = 3$ nuclei, driven by the spin-flavor dependence of the $NN$ interaction {\em and} by the different quark content of the proton and neutron (cf. Fig. 6). In case of $A=3$ systems it should be emphasized that what matters ultimately is the matter size and not the charge radius. The relation between charge and matter radii for $^3He$ and $^3H$ targets differs from Eq. (\ref{eq:radiiA}), namely:
 \be
       <r^2_{c, ^3He}> &  = &  <r^2_{matter, proton}> +  <r^2_{c, proton}> +
       <r^2_{c, neutron}> / 2 \nonumber \\
       <r^2_{c, ^3H}>  &  = &  <r^2_{matter, neutron}> + <r^2_{c, proton}> +
       ~ 2 <r^2_{c,neutron}>, 
       \label{eq:radii3}
 \ee
where $<r^2_{matter, proton}>$ and $<r^2_{matter, neutron}>$ are matter density of proton and neutron in $^3He$. Thus a relation similar to Eq. (\ref{eq:radiiA}) can be obtained only for the isosinglet combination of the $^3He$ and $^3H$ targets since in this case $<r^2_{c, (^3He+^3H) / 2}>  = <r^2_{matter, (proton + neutron) / 2}> + <r^2_{c, proton}> + ~ 5 <r^2_{c, neutron}> / 4$, which coincides with Eq. (\ref{eq:radiiA}) up to the small term $<r^2_{c, neutron}> / 4$.

\indent In case of $^3He$ and $^3H$ nuclei we need to account for the fact that it is the difference between matter radii of proton and neutron in $^3He$ (or in $^3H$) that should be considered in estimating the different $EMC$ effects for these nuclei within the density model. To be able to use the results of the fit of Ref. \cite{Gomez} one should use for proton (neutron) density $\rho_{p(n)}(A) = 3 A / 4 \pi R^3_{p(n)}$, where $R_{p(n)} = \sqrt{5/3} \cdot r_{p(n)}$ and  $r^2_{p(n)} =  <r^2_{matter, proton (neutron)}>  + ~ r_0^2$, where the parameter $r_0$ accounts for the fact that the fit of Ref. \cite{Gomez} is based on the use of the nuclear charge radius. For estimation purposes we take $r_0 \simeq \sqrt{ <r_{c, proton}^2> + <r_{c, neutron}^2>} \simeq 0.7 ~ fm$.

\indent Using these densities one can now estimate the EMC effects within the density model, modifying the Eq. (\ref{eq:density}) as follows:
 \be
        R_{EMC}^{^3He} & \approx & \alpha ~ \left( 1 + \beta ~ {2 \rho_p 
        F_2^p(x, Q^2) + \rho_n F_2^n(x, Q^2) \over 2 F_2^p(x, Q^2) + 
        F_2^n(x, Q^2)} \right) \nonumber \\
        R_{EMC}^{^3H} & \approx & \alpha ~ \left( 1  + \beta ~ {2 \rho_p 
        F_2^n(x, Q^2)  + \rho_n F_2^p(x, Q^2) \over 2 F_2^n(x, Q^2) + 
        F_2^p(x, Q^2)} \right)
        \label{eq:density3}
\end{eqnarray}

\indent In Fig. 11 the predictions of the density model (\ref{eq:density}) for both the $EMC$ ratio and the super-ratio are reported and compared with the results of the $VNC$ model.  For proton and neutron matter radii we adopt the values $\sqrt{<r_{matter, proton}^2>}  = 1.75 \pm 0.03 ~ fm$ and $\sqrt{<r_{matter, neutron}^2>}  = 1.55 \pm 0.04 ~ fm$, obtained from the results of Refs. \cite{Martino,Friar,Schiavilla}. Since in this simple model we neglect small effects of the isospin violation which could lead to $<r^2_{matter, proton} \neq (^3He) \neq <r^2_{matter, neutron}>(^3H)$, the above results correspond to: $\rho_p(^3He) = \rho_n(^3H) \simeq 0.050 ~ fm^{-3}$ and $\rho_n(^3He) = \rho_p(^3H) \simeq 0.068 ~ fm^{-3}$.

\begin{figure}[htb]

\centerline{\epsfxsize=16cm \epsfig{file=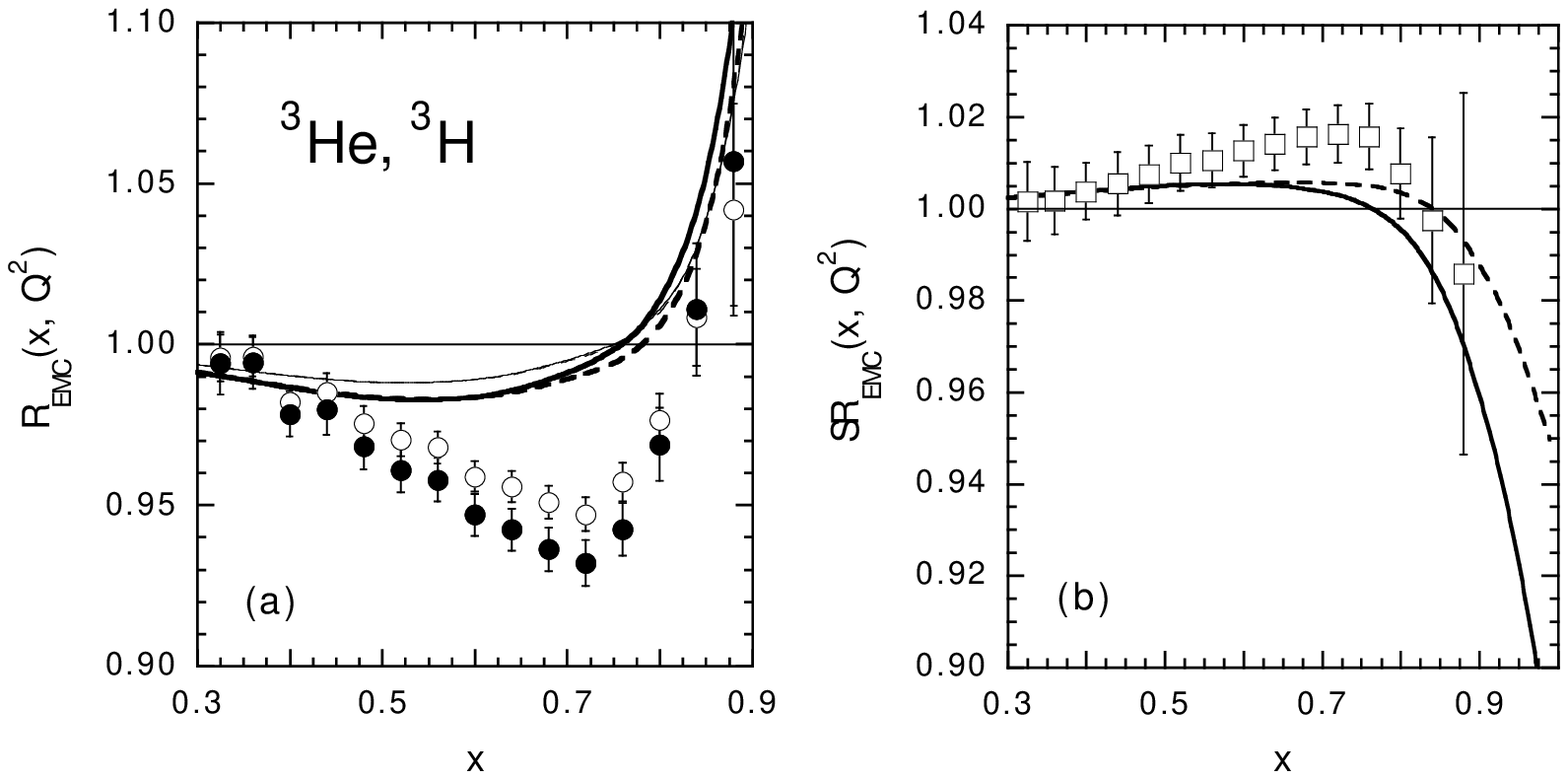}}

{\small {\bf Figure 11.} (a) The $EMC$ ratio [Eq. (\ref{eq:REMC})] in $^3He$ (thin lines) and in  $^3H$ (thick lines) vs. $x$  at $Q^2 = 10 ~ (GeV/c)^2$. (b) The super-ratio [Eq. (\ref{eq:superatio})] vs. $x$. The meaning of the lines is the same as in Fig. 7. In (a) the open and full dots correspond to the predictions of the density model (\ref{eq:density3}) for $^3He$ and $^3H$, respectively, adopting $\rho_p(^3He) = \rho_n(^3H) = 0.050 ~ fm^{-3}$ and $\rho_n(^3He) = \rho_p(^3H) = 0.068 ~ fm^{-3}$. In (b) the prediction of the density model is represented by the open squares.}

\end{figure}

\indent It can be seen from Fig. 11 that the deviation of the $EMC$ ratios from unity is different for $^3He$ and $^3H$ targets by approximately the same amount in percentage, but the density model predicts a deeper $EMC$ effect. Therefore, at variance with the $VNC$ model, the deviation of the super-ratio ${\cal{SR}}_{EMC}(x, Q^2)$ from unity can reach a $\simeq 2 \%$ level already around $x \simeq 0.7 \div 0.8$ in the density model, because the latter predicts a larger $EMC$ effect with respect to the $VNC$ formula.

\indent However we stress again that one should be very careful in applying the density model for light nuclei, as $^3He$ and $^3H$, since for the lightest nuclei the Fermi momentum distribution is very steep and the $A$-dependence of the $EMC$ effect may not have the same form as the one for heavy nuclei. Therefore the predictions of the $EMC$ effect for  $^3He$ and $^3H$ targets based on the density model should be considered for illustrative purposes only.

\subsection{Quark Confinement Size}

\indent In Refs. \cite{Close,Pirner} it was proposed to explain the $EMC$ effect at large $x$ via the softening of the (valence) quark distributions in nuclei (i.e., a more efficient gluon radiation in bound nucleons than in free nucleons) caused by an increase of the confinement volume of the quark in a bound nucleon. One can combine the model of \cite{Close} with the $VNC$ model by including in the latter modifications of the structure functions of the virtual nucleons. For simplicity one can neglect the dependence of the modification on the nucleon momentum treating this effect on average. In this case one can write
 \be
       F_2^A(x, Q^2) = \sum_{N = 1}^A\int_x^A dz ~ z ~ f^N(z)~ F_2^N({x 
       \over z}, \xi_A(Q^2) \cdot Q^2),
       \label{eq:rescaling}
 \ee
where $\xi_A(Q^2)$ is the (dynamical) rescaling factor, whose $Q^2$-dependence, dictated by $pQCD$, is given  by
 \be
      \xi_A(Q^2) = \left[ {\lambda_A^2 \over \lambda_N^2}
      \right]^{\alpha_s(\mu^2) / \alpha_s(Q^2)},
      \label{eq:csiA}
 \ee
with $\lambda_A$ and $\lambda_N$ representing the quark confinement sizes in the bound and free nucleon, respectively. A change of $\lambda_A$ with respect to $\lambda_N$ may be viewed as a change in the nucleon size in the nuclear medium (this interpretation is usually refereed to as the nucleon swelling). In this respect, it should be pointed out that: ~ i) in Ref. \cite{SI85} an increase not larger than $\simeq 6 \%$ of the proton charge radius is found to be compatible with $y$-scaling in $^3He$ and $^{56}Fe$; ~ ii) the analysis of the Coulomb Sum Rule ($CSR$) made in Ref. \cite{CH91} suggests an upper limit of $\simeq 10\%$ for the change of the proton charge radius in $^{56}Fe$; ~ iii) recently \cite{JO96} the experimental values of the $CSR$ in $^{12}C$ and $^{56}Fe$ have been re-analyzed at $Q^2 \simeq 0.3 ~ (GeV/c)^2$, implying an upper limit of $\simeq 8 \%$ for the increase of the proton charge radius (cf. \cite{Ricco}).

\begin{figure}[htb]

\centerline{\epsfxsize=16cm \epsfig{file=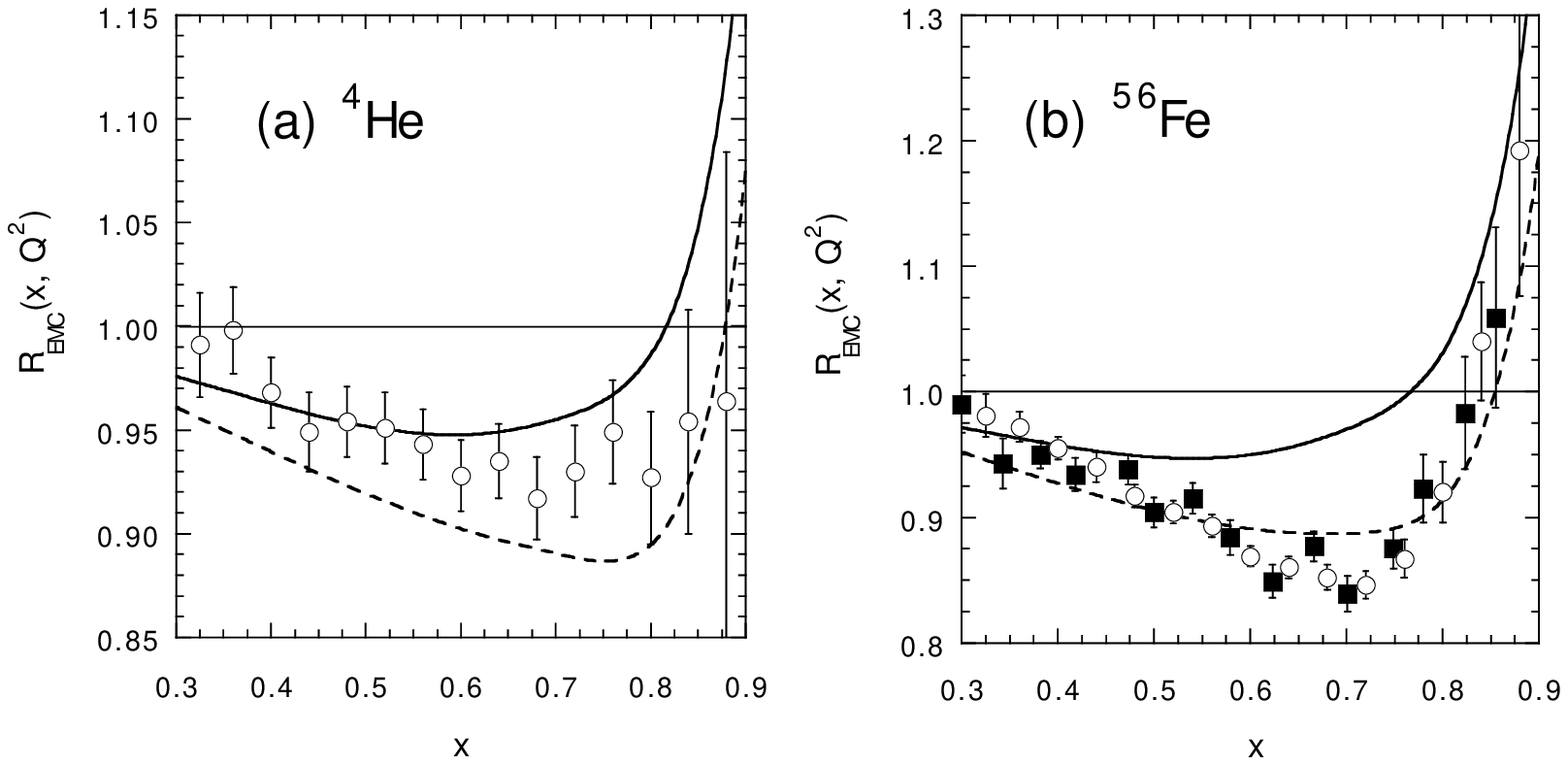}}

{\small {\bf Figure 12.} The $EMC$ ratio [Eq. (\ref{eq:REMC})] in $^4He$ (a) and $^{56}Fe$ (b) at $Q^2 = 10 ~ (GeV/c)^2$. Experimental points and the solid lines are as in Fig. 9. The dashed lines are the results of the rescaling formula (\ref{eq:rescaling}), calculated adopting in Eq. (\ref{eq:csiA}) $\lambda_A / \lambda_N = 1.036$ and $1.047$ for $^4He$ and $^{56}Fe$, respectively.}

\end{figure}

\indent Thus, we assume that ($\lambda_A / \lambda_N - 1$) is proportional to the nuclear density $\rho(A)$ in a such a way that an increase of $6 \%$ is reached only for the heaviest nuclei (namely, at $\rho(A) = 0.15 ~ fm^{-3}$); this corresponds to $\lambda_A / \lambda_N = 1.036$ and $1.047$ for $^4He$ [$\rho(^4He) = 0.089 ~ fm^{-3}$] and $^{56}Fe$ [$\rho(^{56}Fe) = 0.117 ~ fm^{-3}$], respectively. In case of the deuteron we assume no swelling (i.e., $\lambda_D / \lambda_N = 1$). The results of the calculations, adopting for the mass scale $\mu^2$ in Eq. (\ref{eq:csiA}) the value $0.6 ~(GeV/c)^2$ as in \cite{Close,CL}, are reported in Fig. 12 for $^4He$ and $^{56}Fe$ and in Fig. 13 for the mirror $A =3$ nuclei. It can be seen that the $Q^2$-rescaling approach [Eq. (\ref{eq:rescaling})] provides a better description of the $EMC$ data at large $x$ for both $^4He$ and $^{56}Fe$ than the convolution formula (\ref{eq:convolution}). For $^3He$ and $^3H$ the predictions of the rescaling approach (corresponding to $\lambda_{^3He} / \lambda_p = \lambda_{^3H} / \lambda_n =1.020$ and $\lambda_{^3He} / \lambda_n = \lambda_{^3H} / \lambda_p = 1.027$, respectively) provide a possible mechanism to achieve a $\simeq 1.5 \%$ deviation of the super-ratio ${\cal{SR}}_{EMC}(x, Q^2)$ from 1 already around $x \simeq 0.7 \div 0.8$.

\begin{figure}[htb]

\centerline{\epsfxsize=16cm \epsfig{file=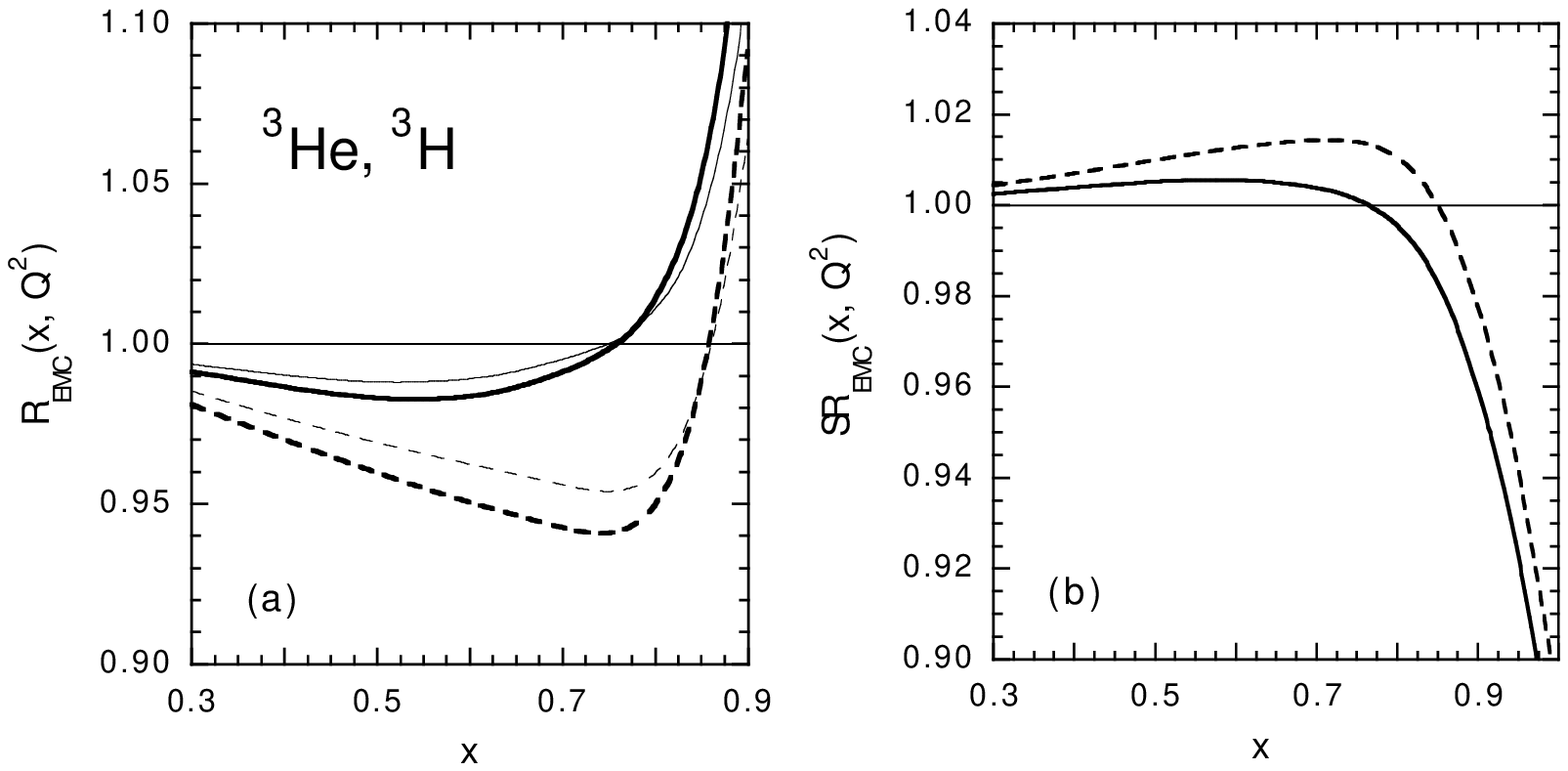}}

{\small {\bf Figure 13.} (a) The $EMC$ ratio [Eq. (\ref{eq:REMC})] in $^3He$ (thin lines) and in $^3H$ (thick lines) vs. $x$  at $Q^2 = 10 ~ (GeV/c)^2$. (b) The super-ratio [Eq. (\ref{eq:superatio})] vs. $x$. The solid and dashed lines correspond to the convolution (\ref{eq:convolution}) and rescaling (\ref{eq:rescaling}) formula, respectively. In the latter case the values $\lambda_{^3He} / \lambda_p = \lambda_{^3H} / \lambda_n =1.020$ and $\lambda_{^3He} / \lambda_n = \lambda_{^3H} / \lambda_p = 1.027$ are adopted in Eq. (\ref{eq:csiA}).}

\end{figure}

\subsection{Six-Quark Clusters}

\indent Another mechanism proposed for the explanation of the $EMC$ effect is the formation of clusters of six (or more) quarks when two (or more) nucleons are overlapping in the nucleus \cite{Vary,Carlson}. This mechanism also provides a softening of the quark distribution in nuclei, since the phase space available in a cluster of six (or more) quarks is clearly larger than in a nucleon. In what follows we limit ourselves to the case of $6q$ clusters and we adopt the procedure of Ref. \cite{CS95} in order to evaluate the $6q$ cluster contribution to the nuclear response.

\indent The main point is to take into account the decomposition (\ref{eq:PpE}) of the nucleon Spectral Function into a ground ($P_0^N$) and a correlated ($P_1^N$) parts. Indeed, since two nucleons can overlap only in the correlated part $P_1^N$, the modification of the convolution formula (\ref{eq:convolution}) due to the possible presence of $6q$ clusters can be written as
 \be
       F_2^A(x, Q^2) & = & \sum_{N = 1}^A \int_x^A dz z f_0^N(z) F_2^N({x 
       \over z}, Q^2) + \sum_{N = 1}^A (1 - {P_{6q} \over S_1^N}) \int_x^A 
       dz z f_1^N(z) F_2^N({x \over z}, Q^2) + \nonumber \\
       & & P_{6q} F_2^{A(6q)}(x, Q^2),
       \label{eq:sixquarks}
 \ee
where, following Eqs. (\ref{eq:fN}) and (\ref{eq:PpE}), one has
 \be
       f_i^N(z) = 2\pi M C_N \int_{E_{min}}^{\infty} dE \int_{p_{min}(z,
       E)}^{\infty} dp ~ p ~ P_i^N(p, E) {M \over \sqrt{M^2 + p^2}},
      \label{eq:f0Nf1N}
 \ee
with $i = 0, 1$. In Eq. (\ref{eq:sixquarks}) $P_{6q}$ is the probability to have a six-quark cluster in the nucleus, $S_1^N$ is the normalization of the correlated part of the nucleon Spectral Function, viz.
 \be
       S_1^N \equiv 4 \pi \int_{E_{min}}^{\infty} dE \int_0^{\infty} dp ~ 
       p^2 ~ P_1^N(p, E) ~,
       \label{eq:S1N}
 \ee
while $F_2^{A(6q)}(x, Q^2)$ is given by
 \be
       F^{A,6q}_2(x, Q^2) = {A \over 2} \sum_{\beta} \int_{{x \over
       2}}^{{M_A \over 2M}} dz_{cm} ~ z_{cm} ~ \tilde{f}^{\beta}(z_{cm}) ~
       \tilde{F}_2^{\beta}({x \over 2z_{cm}}, Q^2),
       \label{eq:F2A6q}
 \ee
where $\beta = (u^2d^4, u^3d^3, u^4d^2) = ([nn], [np], [pp])$ identifies the type of $6q$ cluster, $\tilde{f}^{\beta} (z_{cm})$ is the light-cone momentum distribution describing the center-of-mass motion  of the $6q$ cluster in the nuclear medium and $\tilde{F}_2^{\beta} (\xi, Q^2)$ is the structure function of the $6q$ cluster. Following Ref. \cite{CS95}, we adopt for $\tilde{f}^{\beta} (z_{cm})$ the momentum distribution of the center-of-mass motion of a correlated $NN$ pair (with the same quark content of the $6q$ cluster) as resulting from the Spectral Function model of Ref. \cite{2NC}. In this  way we take into account that the $6q$ bag may be not at rest in the nucleus.

\indent As for $\tilde{F}_2^{\beta} (\xi, Q^2)$ we follow a simple $Q^2$-independent parameterization proposed in Ref. \cite{Carlson}(a) and inspired by quark counting rules, viz.
 \be
       F_2^{\beta}(\xi) = \left\{ a ~ (1 - \xi)^{14} + \left( \sum_j e_j^2 
       \right) b ~ \sqrt{\xi} ~ (1 - \xi)^{10} \right \},
       \label{eq:F26q}
 \ee
where the coefficients $a$ and $b$ can be found in Ref. \cite{Carlson}(a). Note that the charge factor ($\sum_j e_j^2$) is different for the various types of $6q$ clusters, namely: $\sum_j e_j^2 = 4/3, 5/3, 2$ for $[nn], [np], [pp]$ clusters. Therefore, the $6q$ cluster contribution is different in $^3He$ and in $^3H$, because $[nn]$ and $[pp]$ bags have at least different quark content. Note that an additional difference may come from different $x$-distributions in $[pn]$ and $[nn], ~ [pp]$ bags.

\begin{figure}[htb]

\centerline{\epsfxsize=16cm \epsfig{file=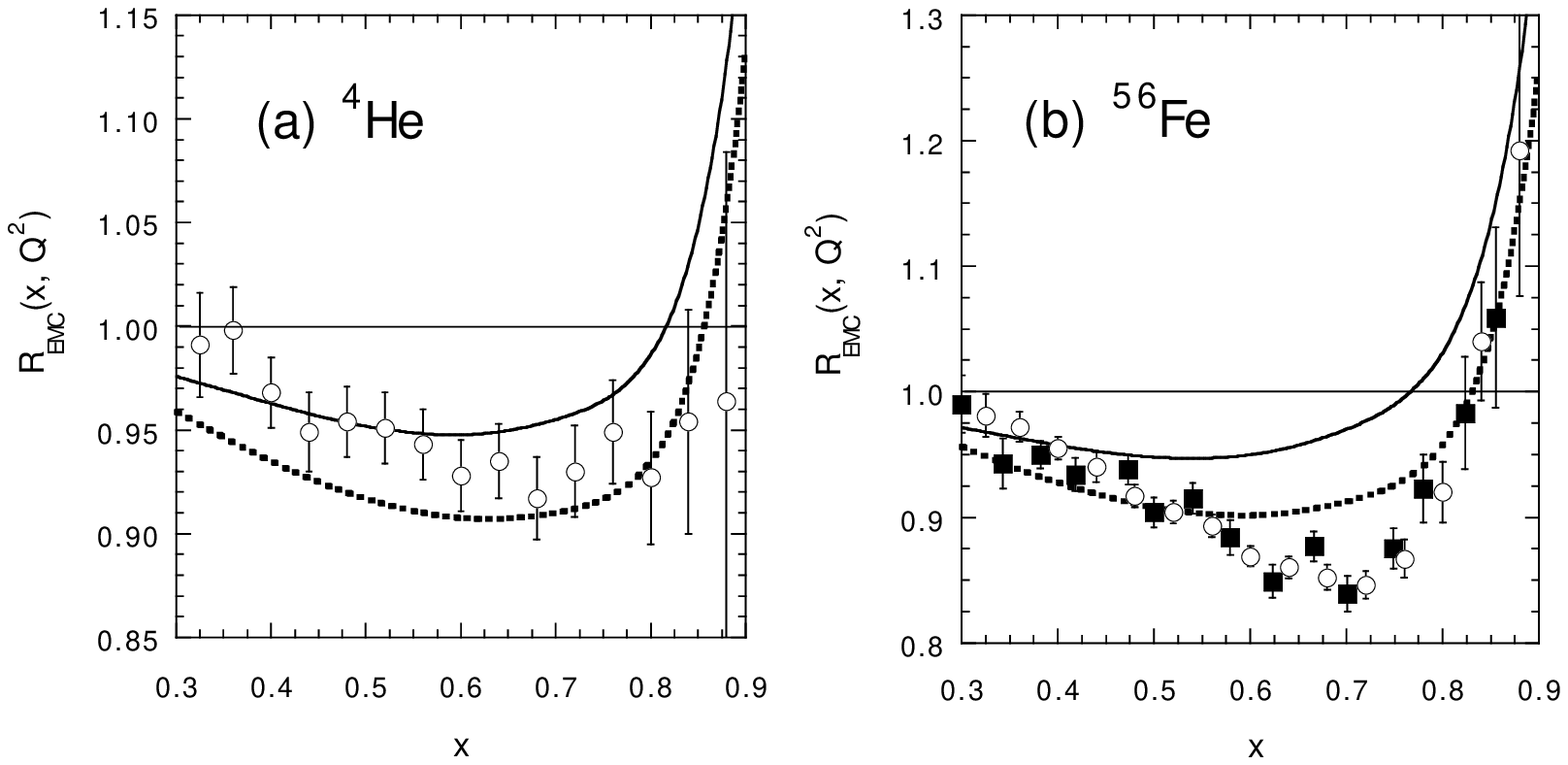}}

{\small {\bf Figure 14.} The $EMC$ ratio [Eq. (\ref{eq:REMC})] in $^4He$ (a) and $^{56}Fe$ (b) at $Q^2 = 10 ~ (GeV/c)^2$. Experimental points and the solid lines are as in Fig. 9. The dotted lines are the results of the $6q$ bag formula (\ref{eq:sixquarks}), calculated adopting $P_{6q} = 11 \%$ and $15 \%$ for $^4He$ and $^{56}Fe$, respectively.}

\end{figure}

\indent The only remaining parameter is the $6q$ bag probability $P_{6q}$ in the nucleus. Since the probability for two nucleon overlapping is proportional to the nuclear density, we assume $P_{6q}$ to be proportional to the density $\rho(A)$. We fix the constant of proportionality by requiring the best reproduction of the $EMC$ data of $^{56}Fe$, obtaining in this way $P_{6q} \simeq 15 \%$ in iron and $P_{6q} \simeq 11 \%$ in $^4He$. We assume no $6q$ bag in the deuteron, while for mirror $A = 3$ nuclei we get $P_{[pp]([nn])} = 6.4 \%$ and $P_{[np]} = 8.6 \%$. The results of our calculations are reported in Figs. 14 and 15. It can be seen that the presence of $6q$ bags can have an important impact on the possible difference of the $EMC$ effect in $^3He$ and $^3H$, leading to a deviation of the super-ratio (\ref{eq:superatio}) of $\simeq 3 \%$ already around $x \simeq 0.7 \div 0.8$.

\begin{figure}[htb]

\centerline{\epsfxsize=16cm \epsfig{file=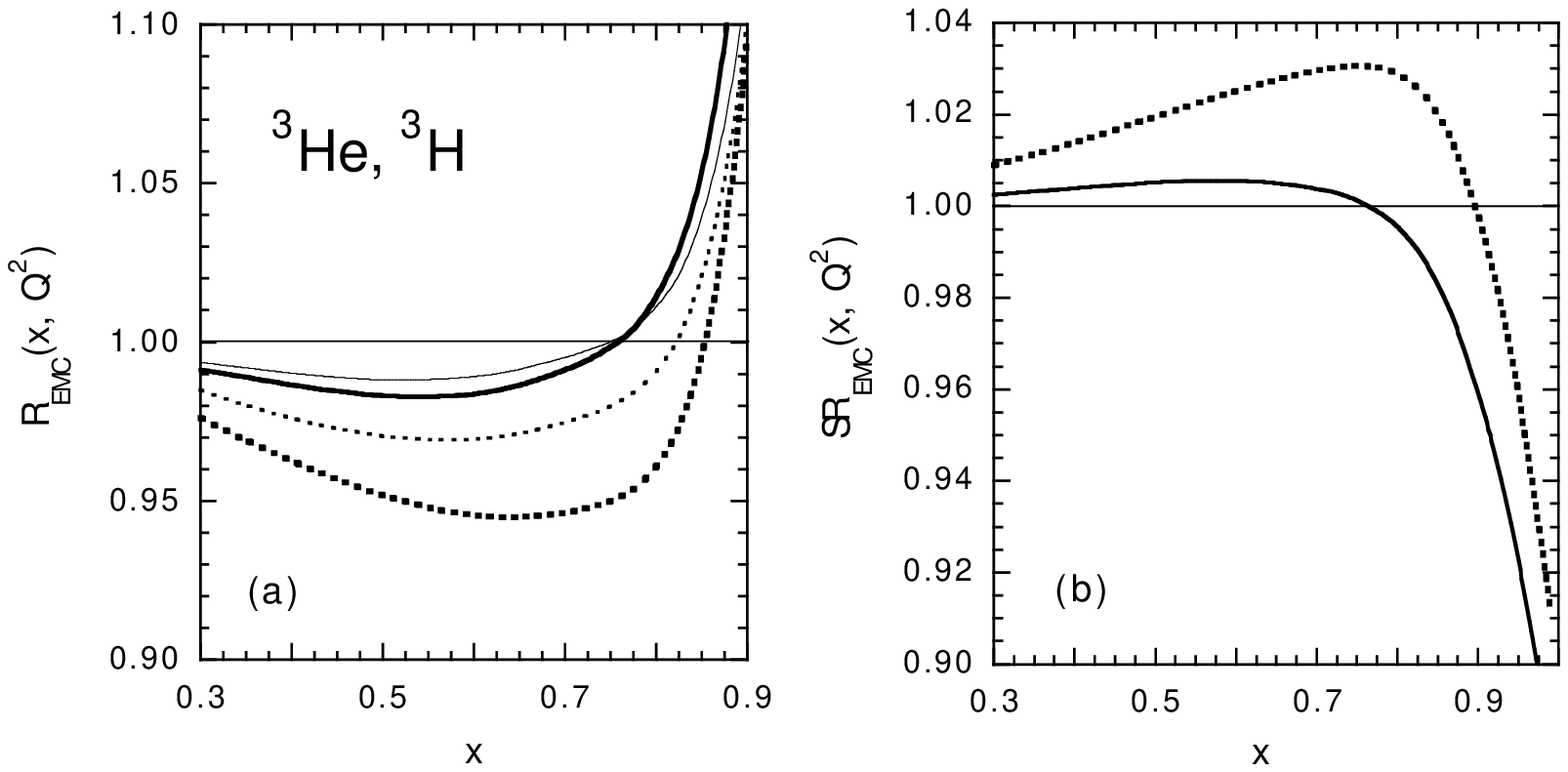}}

{\small {\bf Figure 15.} (a) The $EMC$ ratio [Eq. (\ref{eq:REMC})] in $^3He$ (thin lines) and in $^3H$ (thick lines) vs. $x$  at $Q^2 = 10 ~ (GeV/c)^2$. (b) The super-ratio [Eq. (\ref{eq:superatio})] vs. $x$. The solid and dotted lines correspond to the convolution (\ref{eq:convolution}) and $6q$ bag (\ref{eq:sixquarks}) formula, respectively. In the latter case the values  $P_{[pp]([nn])} = 6.4 \%$ and $P_{[np]} = 8.6 \%$ are adopted.}

\end{figure}

\subsection{Color Screening Model} \label{csm}

\indent In inclusive $A(e, e')X$ reactions the most significant $EMC$ effect is observed at $x \sim 0.5 \div 0.6$. This range of $x$ corresponds to high momentum components of the quark distribution in the nucleon and therefore the $EMC$ effect is expected to be mostly sensitive to nucleon  wave function configurations where three quarks are likely to be close together \cite{FS85,FS}. We refer to such  small size configurations of quarks as point-like configurations ($PLC$). It is then assumed that for large $x$ the dominant contribution to $F_2^N(x,Q^2)$ is given by $PLC$ of partons which, due to color screening, weakly interact with the other nucleons. Note that due to $pQCD$ evolution $F_2^N(x, Q^2)$ at $x \gsim 0.6$, $Q^2 \gsim 10 ~ (GeV/c)^2$, is determined by the non-perturbative nucleon wave function at $x \gsim 0.7$. Thus it is actually assumed that in the non-perturbative nucleon wave function point-like configurations dominate at $x \gsim 0.7$.

\indent The suppression of $PLC$ in a bound nucleon is assumed to be the main source of the $EMC$ effect in inclusive $DIS$ \cite{FS85,FS}. Note that this suppression does not lead to a noticeable change in the average characteristics of the nucleon in nuclei \cite{FS85}. To calculate the change of the probability of a $PLC$ in a bound nucleon,  one can use a perturbation series over a small parameter, $\kappa$,  which controls the corrections to the description of the nucleus as a system of undeformed nucleons. This parameter is taken to be the ratio of the characteristic energies for nucleons and nuclei:
 \be
       \kappa = \left| {\langle U_A \rangle \over \Delta E_A}  \right| \sim 
       {1 \over 10}, 
       \label{plc1}
 \ee
where $\langle U_A \rangle$ is the average potential energy per nucleon, $\langle U_A \rangle \mid_{A \gg 1} \approx - 40 ~ MeV$, and $\Delta E_A \approx M^* - M \sim 0.6 \div 1 ~ GeV$ is the typical energy for nucleon excitations within the nucleus.

\indent The task now is to calculate  the deformation of the quark wave function in the bound nucleon due to suppression of the probability of $PLC$ in a bound nucleon and then to account for it in the calculation of $F_2^A(x, Q^2)$. To this end  we consider a model, in which  the interaction between nucleons is described  by a  Schr\"odinger equation with a potential $V(R_{ij}, y_i, y_j)$ which depends both on the internucleon distances (besides nucleon spin and isospin) and the inner variables $y_i$ and $y_j$, where $y_i$ characterizes the quark-gluon configuration in the $i$-th nucleon \cite{FS85,FS,FJM}.

\indent In the non-relativistic theory of the nucleus the internucleon interaction $V(R_{ij}, y_i, y_j)$ is averaged over all $y_i$ and $y_j$, and the Schr\"odinger equation is solved for the non-relativistic potential $U(R_{ij})$, which is related to $V(R_{ij}, y_i, y_j)$ as follows:
 \be
       U(R_{ij}) = \sum_{y_i, y_j, \widetilde{y}_i, \widetilde{y}_j} 
       \langle \phi_N(y_i) \phi_N(y_j) \mid V(R_{ij}, y_i, y_j, 
       \widetilde{y}_i, \widetilde{y}_j) \mid \phi_N(\widetilde{ y}_i) 
       \phi_N(\widetilde{ y}_j) \rangle, 
       \label{plc3}
   \ee
where $\phi_N(y_i)$ is the free nucleon wave function. Using for the unperturbed nuclear wave function the solution of the Schr\"odinger equation with $U(R_{ij})$, one can treat $(U - V) / (E_i - E_N)$, as a small parameter to calculate the dependence of the probability to find a nucleon in a $PLC$ on the momentum of the nucleon inside the nucleus. The quantity $E_i$ introduced above is the energy of an intermediate excited nucleon state. Such a calculation allows to estimate the suppression of the probability to find a $PLC$ in a bound nucleon as compared to the similar probability for a free nucleon. In the $DIS$ cross section the $PLC$ suppression can be represented as a suppression factor $\gamma_A(p^2)$ which is multiplicative to the nucleon structure function $F_2^N(<\tilde{x}>, Q^2)$ in the $LC$ convolution formula (\ref{F2LC}), viz. \cite{FS85}
 \be
        \gamma_A(p^2) = {1\over (1+\kappa)^2} = {1 \over [1 + (p^2 / M + 
        2\epsilon_A) / \Delta E_A]^2}, \nonumber \\ 
        \label{eq:gamma}
 \ee
where $\Delta E_A = \langle E_i - E_N \rangle \approx M^* - M$ and $p$ is the momentum of the bound nucleon in the light cone.

\indent The $x$ dependence of the suppression effect is based on the assumption that the $PLC$ contribution in the nucleon wave function is negligible at $x \lsim 0.3$, and gives the dominant contribution at $x \gsim 0.5$ \cite{FS85,FSS90}. We use a simple linear fit to describe the $x$ dependence between these two values of $x$ \cite{FSS90}. Using Eq. (\ref{eq:gamma}) for large $A$ at $x \simeq 0.5$ when Fermi motion effects are small one can obtains an estimate for $R_A$ in Eq. (\ref{eq:REMC}) for large  $A$ as follows:
 \be
        R_A(x) \mid_{x \simeq 0.5} \sim \gamma_A(p^2) \approx 1 + {4 \langle
        U_A \rangle \over \Delta E_A} \sim 0.7 \div 0.8 , 
        \label{plct}
 \ee
where $\langle U_A \rangle \approx -40 ~ MeV$. Since $\langle U_A \rangle \sim \langle \rho_A(r)\rangle$ for $A \ge 12$, the model predicts also the $A$ dependence of the $EMC$ effect, which is consistent with the data \cite{FS}. However for the lightest nuclei where the Fermi momentum distribution is very steep, the $A$ dependence due to the nuclear density is rather oversimplified.  The correct estimation requires the convolution of Eq. (\ref{eq:gamma}) with the structure function of a bound nucleon in Eqs. (\ref{F2LC}) and (\ref{F2BLLC}).

\indent To estimate the suppression factor for large Fermi momenta when the interacting nucleon belongs to nucleonic correlations, we use the same formula (\ref{eq:gamma}), in which now $\gamma$ is defined through the virtuality of the interacting nucleon in many-nucleon correlations as follows:
 \be
       \kappa & = & {M_v^2 - M^2 \over M \cdot \Delta E_A}, \nonumber \\ 
       M_v^2 & = & z \left( {M_j^2 \over j} - {M_R^2 + p_{\perp}^2 \over j -
       z} \right),
      \label{kappa}
 \ee
where $M_j \approx j \cdot M$ and $M_R \approx (j - 1) M$ are the masses of the $j$-nucleon correlation and recoil ($j - 1$)-nucleon system.

\begin{figure}[htb]

\centerline{\epsfxsize=16cm \epsfig{file=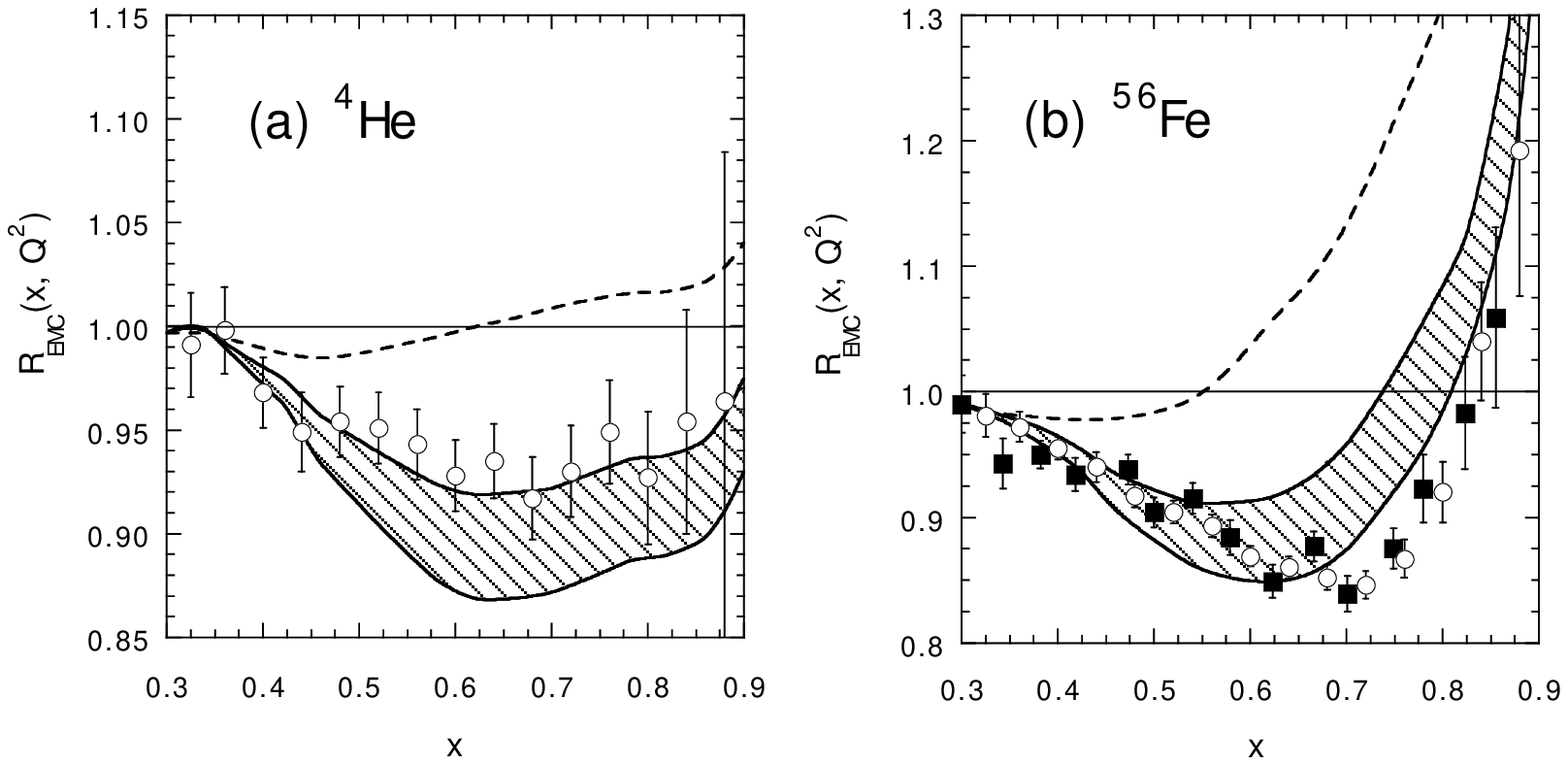}}

{\small {\bf Figure 16.} The $EMC$ ratio in $^4He$ (a) and $^{56}Fe$ (b) at $Q^2 = 10 ~ (GeV/c)^2$. Open dots are data from Ref. \cite{Gomez}, while in (b) the full squares are from Ref. \cite{Arnold}. The dashed lines are the results of Eq. (\ref{F2LC}) calculated adopting the parameterization of the nucleon structure function $F_2^N(x, Q^2)$  of Ref. \cite{Bodek}, including nucleon resonances, and the model of Ref. \cite{2NC} for the nucleon Spectral Function $P^N(k, E)$. The shaded area corresponds to the prediction of the color screening model with $\Delta E = 0.6 ~ GeV$ - lower solid curve and $\Delta E = 1 ~ GeV$ - upper solid curve.}

\end{figure}

\indent In Fig. 16 we present the comparison of the predictions of the color screening model with the $EMC$ data for $^4He$ and $^{56}Fe$ targets. The shaded area is defined by the values $\Delta E = 0.6 ~ GeV$ (lower solid curve) and $\Delta E = 1 ~ GeV$ (upper solid curve).

\indent In Fig. 17 our results obtained for $^3He$ and $^3H$ nuclei assuming the same range of values of $\Delta E$ are presented.  If one assumes that $\Delta E$ is the same for all nuclei, then even though we have large uncertainties for the $EMC$ ratio  [see Figs. 17(a) and 17(b)], the nuclear effects largely cancel out in the super-ratio (\ref{eq:superatio}) [see Fig. 17(c)]. However the assumption that $\Delta E$ is independent on the specific nucleus is a clear oversimplification, since the bound nucleon excitation in $NN$ correlations does depend on the isospin of the $NN$ pair. Indeed, as it is shown in Fig. 10, one has more attraction in isosinglet than in the isovector pairs and  the spatial distributions for different isospin states may substantially differ. Therefore, one can expect that $\Delta E$ is smaller in isotriplet states  as compared to the isosinglet cases. To estimate {\em the upper limit of uncertainties} due to the expected isospin dependence of bound nucleon excitation, we decompose the $NN$ correlation into an isosinglet and an isotriplet contributions. Then we assume for isotriplet states the minimal value of $\Delta E = 0.6 ~ GeV$, whereas for isosinglet states we assume no suppression at all. The prediction of this approximation is shown as the dotted curve in Fig. 17(c).

\begin{figure}[htb]

\centerline{\epsfxsize=16cm \epsfig{file=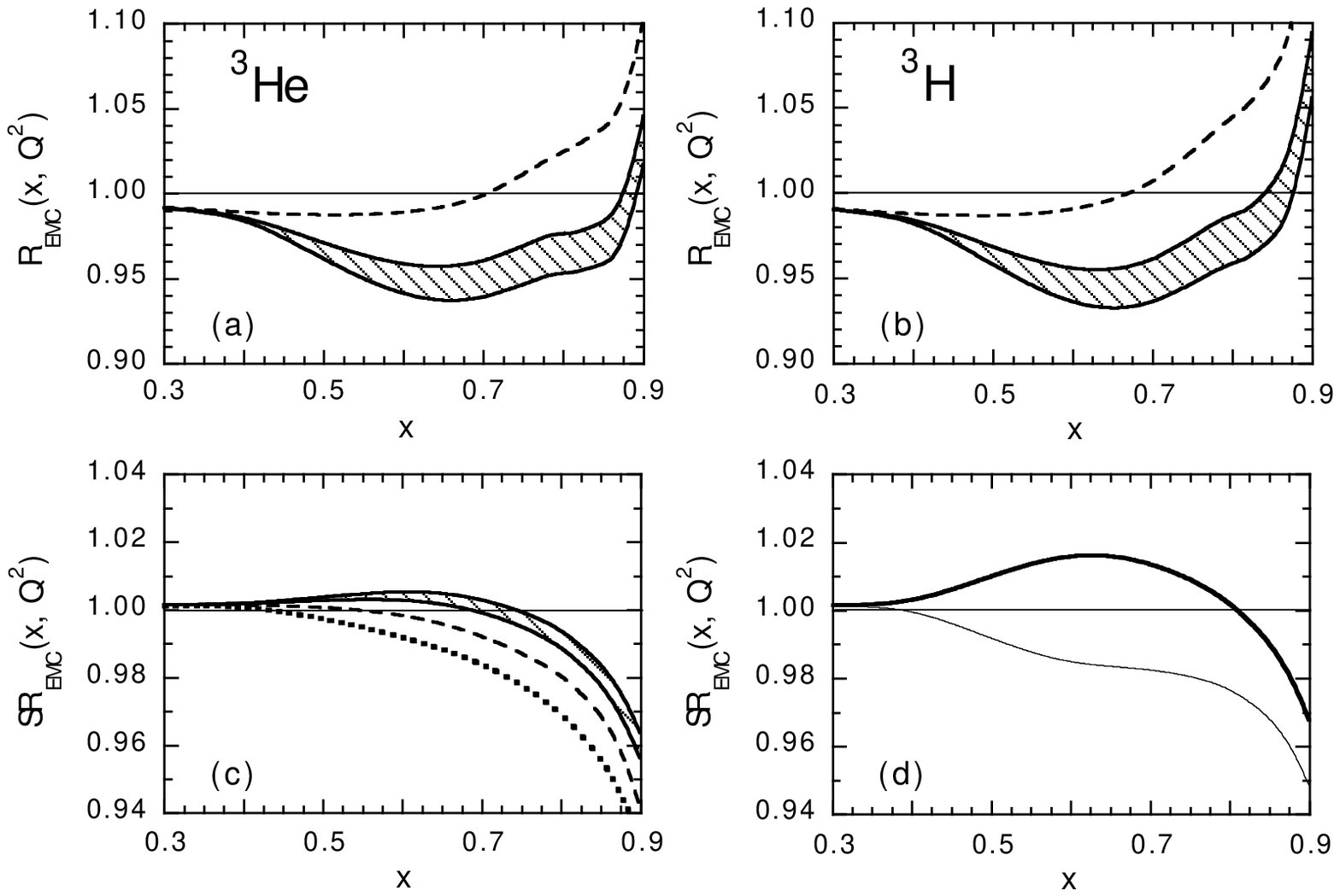}}

{\small {\bf Figure 17.} The $EMC$ ratio in $^3He$ (a) and in $^3H$ (b) vs. $x$ at $Q^2 = 10 ~ (GeV/c)^2$. The super-ratio [Eq. (\ref{eq:superatio})]) vs. $x$ (c, d). In (a)-(c) the dashed curves  correspond to the LC formalism (\ref{F2LC}) using for the nucleon structure function the parameterization of Ref. \cite{Bodek}, which includes nucleon resonances. In (a)-(c) the shaded areas correspond to the predictions of the color screening model for isospin independent values of $\Delta E$, namely for  $\Delta E = 0.6 ~ GeV$ - lower (upper for (c)) solid curves and $\Delta E = 1 ~ GeV$ - upper (lower for (c)) solid curves. In (c) the dotted line corresponds to the prediction of the isospin dependent screening model described in the text. In (d) the thin and thick solid curves correspond to the predictions of the screening model within the quark-diquark picture, when only the valence quarks and the quarks in the diquark are suppressed, respectively.}

\end{figure}

\indent The isospin dependence of the $EMC$ effect emerges naturally also in the quark-diquark model of the nucleon. Within this model one expects a different degree of suppression for valence quark and quarks in the diquark. In Fig. 17(d) we have presented the predictions of the color screening model within  the quark-diquark picture, in which we have considered two extreme cases: the first one when the suppression occurs only for the valence quarks (thin solid curve) and the second one when only the quarks in the diquark are suppressed (thick solid curve). To estimate the extent of the suppression we assume $\Delta E \approx  M_\Delta - M \simeq 0.3 ~ GeV$ which roughly corresponds to the quark helicity-flip excitation in the nucleon.

\indent As follows from Figs. 17(c) and 17(d) all these approximations produce at most a $2 \div 3 \%$ deviation in the super-ratio (\ref{eq:superatio}) around $x \simeq 0.7 \div 0.8$, which may be considered as an upper limit of the uncertainties due to the $EMC$ effect in mirror $A = 3$ nuclei within the color screening model.

\section{Extraction of the $n / p$ Ratio}

\indent We now use the the results of the previous Sections to address the issue of the extraction of the neutron to proton structure function ratio $F_2^n(x, Q^2) / F_2^p(x, Q^2)$ from the measurements of the ratio of the mirror $A = 3$ structure functions, defined as
 \be
       {\cal{R}}_F(x, Q^2) \equiv {F_2^{^3He}(x, Q^2) \over F_2^{^3H}(x, 
       Q^2)} ~. 
       \label{eq:RF}
 \ee
From Eq. (\ref{eq:superatio}) one gets
 \be
       {F_2^n(x, Q^2) \over F_2^p(x, Q^2)} = {2 {\cal{SR}}_{EMC}(x, Q^2) -
       {\cal{R}}_F(x, Q^2) \over 2 {\cal{R}}_F(x, Q^2) - {\cal{SR}}_{EMC}(x,
       Q^2)},
       \label{eq:rationp}
 \ee
and correspondingly the uncertainty on the extracted $n / p$ ratio is given by
 \be
       {\Delta \left( F_2^n / F_2^p \right) \over F_2^n / F_2^p} & = & {3 ~ 
       {\cal{SR}}_{EMC}(x, Q^2) ~ {\cal{R}}_F(x, Q^2) \over [2 
       {\cal{R}}_F(x, Q^2) - {\cal{SR}}_{EMC}(x,  Q^2)] [2 
       {\cal{SR}}_{EMC}(x,  Q^2) - {\cal{R}}_F(x, Q^2)]} \nonumber \\
       & \cdot & \sqrt{ \left[ {\Delta({\cal{SR}}_{EMC}) \over 
       {\cal{SR}}_{EMC}} \right]^2 + \left[ {\Delta({\cal{R}}_F) \over 
       {\cal{R}}_F} \right]^2 }
       \label{eq:delta}
 \ee

\indent One can easily see that in the r.h.s. of Eq. (\ref{eq:delta}) the quantity in front of the square root provides a factor of $\simeq 4$ at $x \gsim 0.7$; thus, even a small uncertainty on the super-ratio is largely amplified in Eq. (\ref{eq:delta}), yielding a non-negligible uncertainty on the extracted $n / p$ ratio at large $x$. Following Ref. \cite{Wally}, the total experimental error in the $DIS$ cross section ratio of $^3H$ and $^3He$ is likely to be $\lsim 1 \%$. Therefore in Eq. (\ref{eq:delta}) we assume that $\Delta({\cal{R}}_F) / {\cal{R}}_F = 1 \%$ and from the results of the previous Sections we consider that $\Delta({\cal{SR}}_{EMC}) / {\cal{SR}}_{EMC} = 3 \%$ for $x \ge 0.6$, leading to a total uncertainty of $\simeq 12 \%$ in the extracted $n / p$ ratio already at $x \simeq 0.7$. In Fig. 18 we have reported the $x$-dependence of the expected accuracy for the extraction of the ratio $F_2^n(x, Q^2) / F_2^p(x, Q^2)$, and the shaded areas include the combination of all the effects discussed above. Moreover we have used the $CTEQ$ \cite{CTEQ} set of $PDF$'s (lower shaded area) and the modified $CTEQ$ parameterization (upper shaded area), obtained from the $CTEQ$ one as described in Sec. 3, in order to reproduce a $n / p$ ratio approaching $3 / 7$ at $x \to 1$. Figure 18 demonstrates that although mirror $A = 3$ measurements will significantly improve the existing accuracy of the neutron structure functions at large $x$, they may not provide $3 \sigma$ separation for the two predictions of the $n / p$ ratio having limiting values of $1 / 4 $ and  $3 / 7$  at $x \to 1$.

\begin{figure}[htb]

\centerline{\epsfxsize=16cm \epsfig{file=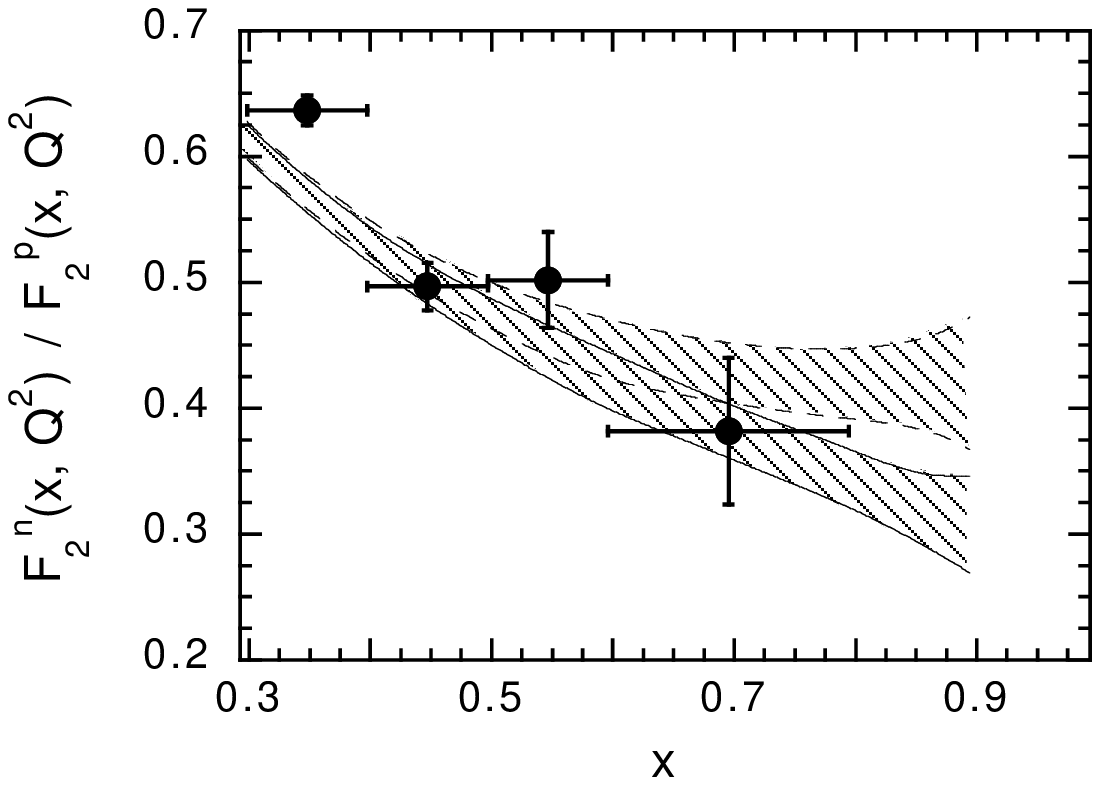}}

{\small {\bf Figure 18.} The expected accuracy for the extraction of the neutron to proton ratio $F_2^n(x, Q^2)/ F_2^p(x, Q^2)$ vs. $x$ at $Q^2 = 10 ~ (GeV/c)^2$. The lower and upper shaded areas correspond to the $CTEQ$ and modified $CTEQ$ parameterization described in the text. The full dots are the $NMC$ data points as given in  Ref. \cite{Arneodo}.

\label{mbexp}}

\end{figure}

\indent In Ref. \cite{Wally,Pace} it was suggested that, once the ratio ${\cal{R}}_F(x, Q^2)$ is measured, one can employ an iterative procedure to extract the $n / p$ ratio which can almost eliminate the effects of the dependence of the super-ratio ${\cal{SR}}_{EMC}(x, Q^2)$ on the large-$x$ behavior of the specific structure function input (see Fig. 6). Namely, after extracting the $n / p$ ratio assuming a particular calculation of ${\cal{SR}}_{EMC}$, one can use the extracted neutron structure function to get a new estimate of ${\cal{SR}}_{EMC}$, which can then be employed for a further extraction of the $n / p$ ratio. Such a procedure can be iterated until convergence is achieved and self-consistent solutions for the extracted $F_2^n / F_2^p$ and the super-ratio ${\cal{SR}}_{EMC}$ are obtained. In Ref. \cite{Pace} the numerical estimate of the iteration procedure was performed within the $VNC$ model and a good convergence was achieved for $x$ up to $\simeq 0.8$. However this result depends on the assumed validity of the $VNC$ model in the considered range of values of $x$. To check how well the iteration procedure will work in case of other models of the $EMC$ effect, we have considered the following two examples.

\indent First, let us consider in Eq. (\ref{eq:rationp}) the nuclear structure function ratio ${\cal{R}}_F(x, Q^2)$ which results from the use of the modified $SLAC$ parameterization of $F_2^N(x, Q^2)$ and the inclusion of the effects of possible $6q$ bags within the $VNC$ model. Then we apply the iteration procedure assuming for the super-ratio ${\cal{SR}}_{EMC}(x, Q^2)$ the convolution calculation corresponding to the Donnachie-Landshoff ($DL$) fit \cite{DL} of $F_2^N(x, Q^2)$, which provides a $n / p$ ratio equal to $1 / 3$ as $x \to 1$. Figure 19 demonstrates that a consistency is achieved between the $n / p$ ratio used as the input and the extracted one. However, for the calculation of ${\cal{R}}_F(x, Q^2)$ we started from the modified $SLAC$ parameterization which goes to $3 / 7$ as $x \to 1$.

\begin{figure}[htb]

\centerline{\epsfxsize=12cm \epsfig{file=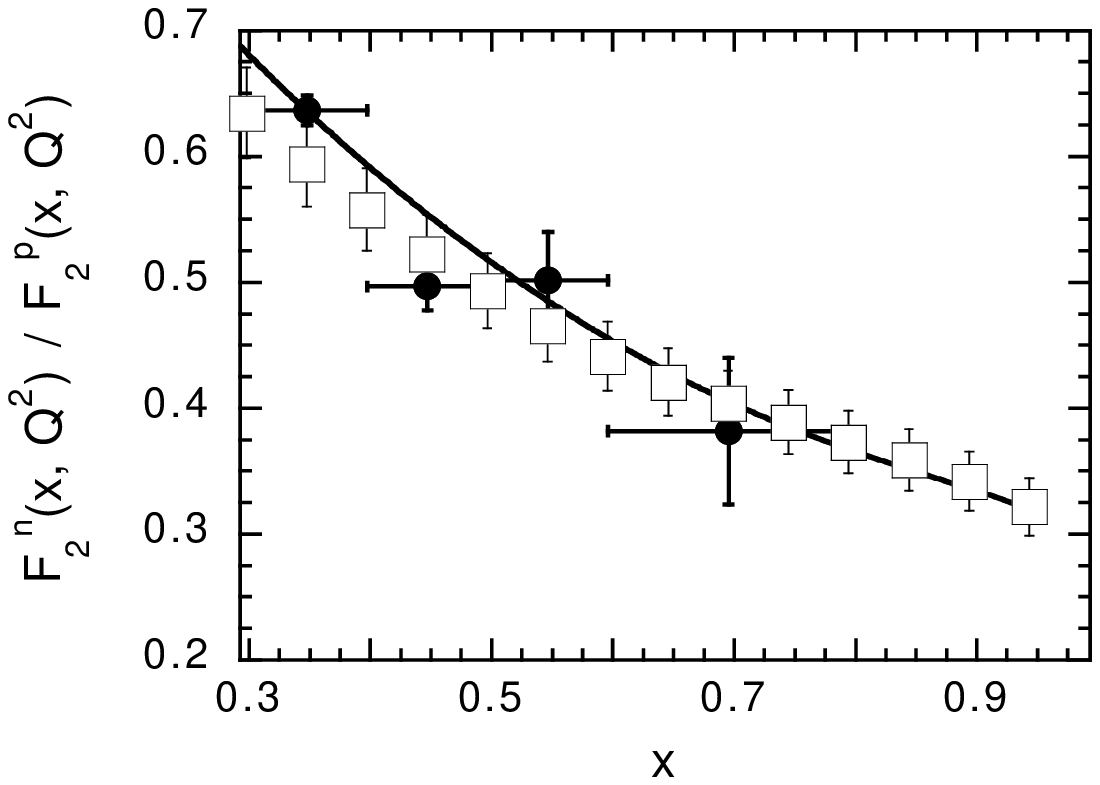}}

{\small {\bf Figure 19.}  The neutron to proton ratio $F_2^n(x, Q^2) / F_2^p(x, Q^2)$ vs. $x$ at $Q^2 = 10 ~ (GeV/c)^2$. The full dots are the $NMC$ data points as given in Ref. \cite{Arneodo}. The solid line represents the $DL$ parameterization \cite{DL} of the $n / p$ ratio. The open squares are the results of the extraction of the $n / p$ ratio adopting the  the convolution formula (\ref{eq:convolution}) as described in the text. The error bars are calculated via Eq. (\ref{eq:delta}).}

\end{figure}

\indent In the second example we calculate the nuclear structure function ratio ${\cal{R}}_F(x, Q^2)$ using the modified $CTEQ$ parameterization within the $LC$ approximation adopting the color screening model for the $EMC$ effect (only with valence quark suppression in the quark-diquark picture). To do the iteration we start with the super-ratio ${\cal{SR}}_{EMC}(x, Q^2)$ calculated within the $LC$ approximation without $EMC$ effects using the $CTEQ$ parameterization for the nucleon structure function. Figure 20 demonstrates that the iteration diverges already at values of $x$ ($\simeq 0.7$) smaller than the ones obtained in \cite{Pace}, where only the $VNC$ model was used. Below $x \simeq  0.7$ the iteration procedure converges to a value of  the $n / p$ ratio which differs from the ``exact'' one (used in $R_F(x, Q^2)$) exactly by the amount of the $EMC$ effect which is implemented in the calculation of $R_F(x, Q^2)$.

\begin{figure}[htb]

\centerline{\epsfxsize=12cm \epsfig{file=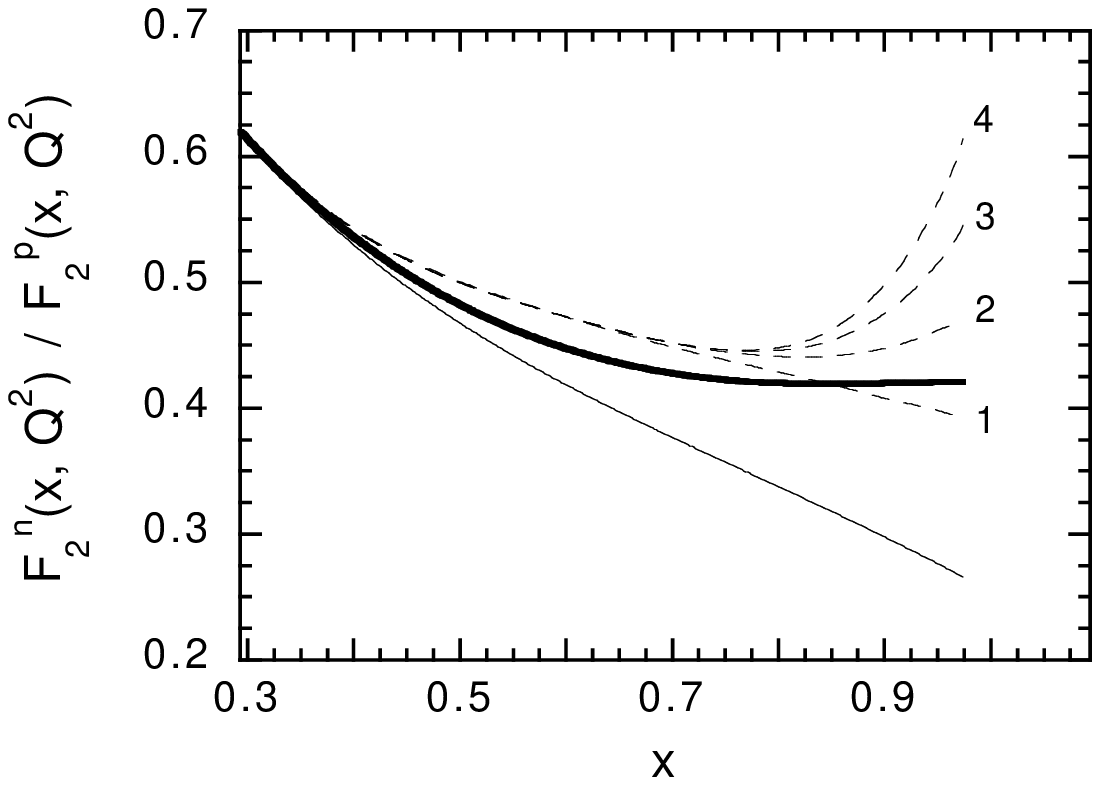}}

{\small {\bf Figure 20.} The $x$ dependence of neutron to proton structure function ratios-$F_{2n}/F_{2p}$. The thick solid curve corresponds to the $n / p$ ratio used to calculate $R_F$. The thin solid curve is the ratio used as input for the iteration procedure. The dashed curves correspond to the extracted ratio obtained after subsequent iterations.}

\label{iter}

\end{figure}

\indent Both examples  illustrate  that the iterative procedure cannot improve the accuracy of the extraction of the $n / p$ ratio as estimated in Fig. 18.

\section{Conclusions}

\indent In conclusion, we have performed a detailed theoretical investigation of the $EMC$ effect in $^3He$ and $^3H$ targets. First, we have considered first the convolution approximation within the $VNC$ model and the $LC$ approach. The differences in the motion of protons and neutrons in mirror $A = 3$ nuclei, resulting from the spin-flavor dependence of the nuclear force, have been taken into account. We have investigated various sources of uncertainties in the estimate of the super-ratio of the $EMC$ effects in mirror $A = 3$ nuclei, like: ~ i) charge symmetry breaking terms in the nucleon-nucleon interaction; ~ ii) finite $Q^2$ corrections to the Bjorken limit; ~ iii) the role of different prescriptions to relate the invariant nucleon Spectral Function to the non-relativistic one, required to ensure baryon number conservation; ~ iv) the role of the $x$-shape of parton distribution functions; and ~ v) the differences between the $VNC$ model and the light cone formalism.

\indent Within convolution approach, in which no modification of the bound nucleon is considered, the deviation of the super-ratio (\ref{eq:superatio}) from unity is predicted to stay within $1 \%$ only for $x \lsim 0.75$ in close agreement with Ref. \cite{Pace} and in overall agreement with Ref. \cite{Wally} (which by the way neglect the effects (i)-(iv)), where $1 \%$ deviations were found from an average value of the super-ratio equal to $\approx 1.01$.

\indent We have further argued that the previous estimate cannot be considered as definitive for the purpose of extraction of the $F_2^n/F_2^p$ ratio, since it is derived using just one of the many models of the $EMC$ effect, which has in particular a number of problems in describing the nuclear data, namely the underestimation the $EMC$ effect at $x \gsim 0.6$. Hence we have provided a detailed analysis of the super-ratio within a broad range of models of the $EMC$ effect, which take into account possible modifications of the bound nucleons in nuclei, like: ~ i) a change in the quark confinement size (including swelling); ~ ii) the possible presence of clusters of six quarks; and ~ iii) the suppression of point-like configurations due to color screening. Our main result is that one cannot exclude the possibility that the cancellation of the nuclear effects in the super-ratio may occur only at  a level of $\approx 3 \%$, resulting in a significant uncertainty (up to $\approx 12 \%$ for $x \approx 0.7 \div 0.8$) in the extraction of the free $n / p$ ratio from the ratio of the measurements of the $^3He$ and $^3H$ $DIS$ structure functions. Such an uncertainty is comparable to the $\simeq 18 \%$ difference between the $n / p$ predictions having limiting values of $3 / 7$ and $1/4$ at $x \to 1$, which characterize the $pQCD$ and Feynman models. Another consequence of the use of a broad range of models for the $EMC$ effect is that the iteration procedure cannot in general improve the accuracy of the extracted $n / p$ ratio.

\indent It is however important to note that despite of such restrictions the mirror $A = 3$ measurements will provide an unprecedented accuracy in the extraction of the neutron $DIS$ structure function. Thus such measurements are very much welcomed.  It is however very important to complement them with the measurements of semi-inclusive processes off the deuteron, in  which the momentum of the struck nucleon is tagged by detecting the recoiling one. Imposing the kinematical conditions that the detected momentum is low ($p \lsim 150 ~ MeV/c$), which means that the nucleons in the deuteron are initially far apart \cite{FS,Strikman,SIM96}, it is possible to minimize significantly the nuclear effects. Furthermore, all the unwanted nuclear effects can be isolated by using the same reaction for the extraction of the proton structure function by detecting slow recoiling neutrons and comparing the results with existing hydrogen data, as well as by performing tighter cuts on the momentum of the spectator proton and then extrapolating to the neutron pole.

\section*{Acknowledgements}

\indent The authors gratefully acknowledge Steven Pieper for supplying them with the results of the Green Function Monte Carlo method of Ref. \cite{Pieper}. M.M.S. gratefully acknowledges a contract from Jefferson Lab under which this work was done. The Thomas Jefferson National Accelerator Facility (Jefferson Lab) is operated by the Southeastern Universities Research Association (SURA) under DOE contract DE-AC05-84ER40150. This work is supported also by DOE grants under contract DE-FG02-93ER-40771 and DE-FG02-01ER41172. We  are grateful to the Institute for Nuclear Theory (Seattle) for the hospitality during the course of this study.


\begin{thebibliography}{99}

\bibitem{Yang} U.K. Yang and A. Bodek: Phys. Rev. Lett. {\bf 82}, 2467 
 (1999).

\bibitem{Feynman} R.P. Feynman: {\em Photon Hadron Interactions}, Benjamin 
 Inc. 1972

\bibitem{Isgur} For a recent review see N. Isgur: Phys. Rev. {\bf D59}, 
 034013 (1999) and references therein quoted

\bibitem{Brodsky} S.J. Brodsky and G. Farrar: Phys. Rev. {\bf D11}, 229 
 (1975).

\bibitem{pQCD} G.R. Farrar and D.R. Jackson: Phys. Rev. Lett. {\bf 35}, 1416
 (1975).

\bibitem{Bodek} (a) A. Bodek et al.: Phys. Rev. {\bf D20}, 1471 (1979).  (b)
 A. Bodek and J.L. Ritchie: Phys. Rev. {\bf D23}, 1070 (1981).

\bibitem{West} G.B. West: Phys.Lett. {\bf  37B}, 509 (1971); Ann. Phys. 
 (N.Y.) {\bf 74}, 464 (1972). W.B. Atwood and G.B. West:  Phys. Rev.  {\bf 
 D7}, 773 (1973).

\bibitem{DATA} J.J. Aubert et al.: Nucl. Phys. {\bf B293}, 740 (1987). A.C. 
 Benvenuti et al.: Phys. Lett. {\bf B237}, 599 (1990). P. Amaudruz et al.: 
 Nucl. Phys. {\bf B371}, 3 (1992).

\bibitem{SLAC} L.W. Whitlow et al.: Phys. Lett. {\bf 282B}, 475 (1992).

\bibitem{CTEQ} H.L. Lai et al.: Phys. Rev. {\bf D51}, 4763 (1995).

\bibitem{GRV} M. Gl\"uck, E. Reya and A. Vogt: Z. Phys. {\bf C67}, 433 
 (1995).

\bibitem{FS76west} L.L. Frankfurt and M.I. Strikman: Phys. Lett. {\bf B64}, 
 433 (1976); Phys. Lett. {\bf B76}, 333 (1978).

\bibitem{LP78} P.V. Landshoff and J.C. Polkinghorne: Phys. Rev.  {\bf D18}, 
 153 (1978).

\bibitem{FS76lc} L.L. Frankfurt and M.I. Strikman: Phys. Lett.  {\bf B65}, 
 51 (1976).

\bibitem{EMCexp}J.J.~Aubert {\em et al.} (EMC collaboration): Phys. Lett. 
 {\bf B 123}, 275 (1983).

\bibitem{FS85} L.L. Frankfurt and M.I. Strikman: Nucl. Phys. {\bf B250} 
 (1985) 143.

\bibitem{FS} L.L. Frankfurt and M.I. Strikman: Phys. Rep. {\bf 160}, 235 
 (1988).

\bibitem{S92}M.I. Strikman, {\em Nuclear Parton Distributions and Extraction 
 of Neutron Structure Functions}, in Proc. of XXVI International Conference 
 on High Energy Physics, Dallas (USA),  World Scientific, Singapore, Vol. 1,
 8060809 (1992).

\bibitem{BDR}A. Bodek, S. Dasu and S.E. Rock, in Tucson Part. Nucl. Phys. 
 768-770 (1991).

\bibitem{OFF-SHELL} L.P. Kaptari and A.Yu. Umnikov: Phys. Lett. {\bf 
 B259},155 (1991). M.A. Braun and M.V. Tokarev: Phys. Lett. {\bf B320}, 381 
 (1994). A.Yu. Umnikov and F.C. Khanna: Phys. Rev. {\bf C49}, 2311 (1994). 
 W. Melnitchouk, A.W. Schreiber and A.W. Thomas: Phys. Rev. {\bf D49}, 1183 
 (1994); Phys. Lett. {\bf 335B}, 11 (1994). S.A. Kulagin, G. Piller and W. 
 Weise: Phys. Rev. {\bf C50}, 1154 (1994).

\bibitem{MT96} W. Melnitchouk and A.W. Thomas: Phys. Lett. {\bf B377}, 11 
 (1996).

\bibitem{Ricco_NPB} G. Ricco, S. Simula and M. Battaglieri: Nucl. Phys. {\bf
 B555}, 306 (1999); Nucl. Phys. {\bf A663}, 1015 (2000); {\em ibidem} {\bf 
 A666}, 165 (2000).

\bibitem{Strikman}W. Melnitchouk, M. Sargsian and M.I. Strikman:  in Proc. 
 of the Workshop on "Future Physics at HERA" (DESY, September '95 - May 
 '96), ed. by G. Ingelman, A. de Roeck and R.K. Klanner, DESY (Hamburg, 
 1996), p. 1064,  also e-print archive nucl-th/9609048; Z. Phys. {\bf A359}, 
 99 (1997)

\bibitem{SIM96} S. Simula: Phys. Lett. {\bf B387}, 245 (1996); Nucl. Phys. 
 {\bf A631}, 602c (1997); in Proc. of the Workshop on "Future Physics at 
 HERA" (DESY, September '95 - May '96), ed. by G. Ingelman, A. de Roeck and 
 R.K. Klanner, DESY (Hamburg, 1996), p. 1058, also e-print archive nucl-th 
 9608053.

\bibitem{Wally} I.R. Afnan et al.: Phys. Lett. {\bf B493}, 36 (2000).

\bibitem{Pace} E. Pace, G. Salm\'e and S. Scopetta: Phys. Rev. {\bf C64}, 
 05503 (2001).

\bibitem{Chew} G.F. Chew and F.E. Low: Phys. Rev. {\bf 113}, 1640 (1959).

\bibitem{DY} D.M. Alde et al.: Phys. Rev. Lett. {\bf 64}, 2479 (1990).

\bibitem{g1} L. Frankfurt, V. Guzey and M. Strikman:  Phys. Lett. {\bf 
 B381}, 379 (1996). F. Bissey, V. Guzey, M. Strikman and A.W. Thomas: 
 e-print archive hep-ph/0109069.

\bibitem{FS87} L.L. Frankfurt and M.I. Strikman: Phys. Lett. {\bf B183}, 254
 (1987).

\bibitem{FSS90} M.M. Sargsian, L.L. Frankfurt and M.I. Strikman: Z. Phys. 
 {\bf A335}, 431 (1990).

\bibitem{FS81} L.L. Frankfurt and M.I. Strikman: Phys. Rep. {\bf 76}, 215 
 (1981).

\bibitem{2NC} (a) C. Ciofi degli Atti, L.L. Frankfurt, S. Simula and M.I. 
 Strikman: Phys. Rev. {\bf C44}, R7 (1991). (b) C. Ciofi degli Atti and S. 
 Simula: Phys. Rev. {\bf C53}, 1689 (1996).

\bibitem{EMC} E.L. Berger and F. Coester: Ann. Rev. Part. Sci. {\bf 37}, 463 
 (1987). T. Uchiyama and K. Saito: Phys. Rev. {\bf C38}, 2245 (1988). R.P. 
 Bickerstaff and A.W. Thomas: J. Phys. {\bf G15}, 1523 (1989).

\bibitem{CL} C. Ciofi degli Atti and S. Liuti: Phys. Lett. {\bf B225}, 215 
 (1989); Phys. Rev. {\bf C41}, 1100 (1990); Phys. Rev. {\bf C44}, 1269 
 (1991).

\bibitem{Sauer} H. Meier-Hajduk, P.U. Sauer and W. Theis: Nucl. Phys. {\bf 
 A395}, 332 (1983).

\bibitem{CPS} C. Ciofi degli Atti, E. Pace and G. Salm\'e: Phys. Rev. {\bf 
 C21}, 805 (1980).

\bibitem{BFF}O. Benhar, A. Fabrocini and S. Fantoni: Nucl. Phys. {\bf A505},
 267 (1989); {\bf A550}, 201 (1992).

\bibitem{propagator} A. Ramos, A. Polls and W.H. Dickhoff: Nucl. Phys. {\bf 
 A503}, 1 (1989). W.H. Dickhoff and H. Muther: Rep. Prog. Phys. {\bf 55}, 
 1947 (1992). H.S. Kohler: Nucl. Phys. {\bf A537}, 64 (1992); Phys. Rev. 
 {\bf C46}, 1687 (1992). M. Baldo et al.: Nucl. Phys. {\bf A545}, 741 
 (1992).  F. de Jong and R. Malfliet: Phys. Rev. {\bf C44}, 998 (1991). P. 
 Fern\'andez de C\'ordoba and E. Oset: Phys. Rev. {\bf C46}, 1697 (1992).

\bibitem{RSC} R.V. Reid: Ann. Phys. (N.Y.) {\bf 50}, 411 (1968).

\bibitem{Nogga} A. Nogga, H. Kamada and W. Glockle: Phys. Rev. Lett. {\bf 
 85}, 944 (2000) and references therein quoted.

\bibitem{Pieper} S.C. Pieper and R.B. Wiringa: Annu. Rev. Nucl. Part. Sci. 
 {\bf 51}, 53 (2001).

\bibitem{Day} D. Day et al.: Phys. Rev. Lett. {\bf 43} (1979) 1143.

\bibitem{QE} L.L  Frankfurt, M.M. Sargsian and M.I. Strikman: in 
 preparation.

\bibitem{Arneodo} M. Arneodo et al.: Phys. Rev. {\bf D50}), R1 (1994).

\bibitem{Gomez} J. Gomez et al.: Phys. Rev. {\bf D49}, 4348 (1994).

\bibitem{Arnold} R.G. Arnold et al.: Phys. Rev. Lett. {\bf 52}, 727 (1984).

\bibitem{Jaffe} R.L. Jaffe, F.E. Close, R.G. Roberts and G.G. Ross: Phys. 
 Lett. {\bf B134}, 449 (1984).

\bibitem{Close} F.E. Close, R.G. Roberts and G.G. Ross: Phys. Lett. {\bf 
 B129}, 346 (1983).  F.E. Close et al.: Phys. Rev. {\bf D31}, 1004 (1985).

\bibitem{Pirner} O. Nachtmann and H.J. Pirner: Z. Phys. {\bf C21}, 277 
 (1984). G. G\"uttner and H.J. Pirner: Nucl. Phys. {\bf 457}, 555 (1986).

\bibitem{Vary} J. Vary et al.: Phys. Rev. Lett. {\bf 46}, 1376 (1981).

\bibitem{Carlson} (a) C.E. Carlson and T.J. Havens: Phys. Rev. Lett. {\bf 
 51}, 261 (1983). (b) K.E. Lassila and U.P. Sukhatme: Phys. Lett. {\bf 
 B209}, 343 (1988).

\bibitem{FJM} M.R. Frank, B.K. Jennings and G.A. Miller: Phys. Rev. {\bf 
 C54}, 920 (1996).

\bibitem{Berger} E.L. Berger, F. Coester and R.B. Wiringa: Phys. Rev. {\bf 
 D29}, 398 (1984). E.L. Berger and F. Coester: Phys. Rev. {\bf D32}, 1071 
 (1985).

\bibitem{Martino} F.-P. Juster et al.: Phys. Rev. Lett. {\bf 55}, 2261 
 (1985). J. Martino: Lecture Notes in Physics {\bf 260}, Springer Verlag 
 (1986), p. 129.

\bibitem{Friar} C.R. Chen et al.: Phys. Rev. {\bf C31}, 2266 (1985); {\em 
 ibidem} {\bf C33}, 1740 (1986). J.L. Friar et al.: Phys. Rev. {\bf C35} , 
 1502 (1987).

\bibitem{Schiavilla} R. Schiavilla, V.R. Pandharipande and R.B. Wiringa: 
 Nucl. Phys. {\bf A449}, 219 (1986).

\bibitem{SI85} I. Sick: Phys. Lett. {\bf B157}, 13 (1985).

\bibitem{CH91} J.P. Chen et al.: Phys. Rev. Lett. {\bf 66}, 1283 (1991).

\bibitem{JO96} J. Jourdan: Nucl. Phys. {\bf A603}, 117 (1996).

\bibitem{Ricco} G. Ricco et al.: Phys. Rev. {\bf C57}, 356 (1998); Few Body 
 Syst. Suppl. {\bf 10}, 423 (1999).

\bibitem{CS95} (a) C. Ciofi degli Atti and S. Simula: Few Body Systems {\bf 
 18}, 55 (1995). (b) S. Simula: Few Body Systems Suppl. {\bf 8}, 423 (1995); 
 {\em ibidem} {\bf 9}, 466 (1995).

\bibitem{DL} A. Donnachie and P.V. Landshoff: Zeit Phys. {\bf C61}, 139 
 (1994).

\end{thebibliography}
\end{document}